\documentclass[journal]{IEEEtran}
\IEEEoverridecommandlockouts
\usepackage[table]{xcolor}
\usepackage{cite}
\usepackage{amsmath,amssymb,amsfonts}
\usepackage{graphicx}
\usepackage{textcomp}
\usepackage{xcolor}
\usepackage{algorithm}
\usepackage{tabularx}
\usepackage{algpseudocode}
\usepackage{pifont}
\usepackage{placeins}
\usepackage{mathtools}

\usepackage{subfig}
\usepackage{bbold}
\usepackage[nocomma]{optidef}

\usepackage{setspace}
\newcommand{\subparagraph}{}
\DeclareMathOperator*{\argmax}{arg\,max}

\DeclareMathOperator*{\diag}{diag}

\begin{document}

% Keywords command
\providecommand{\keywords}[1]
{
  \textbf{\textit{Keywords---}} #1
}

	\title{
	%Enabling content delivery for Internet of Vehicles using Proximal Policy  Optimization	
	%Joint trajectory planning and cache management for content delivery to internet of vehicles
	%Caching-Empowered RSU-Assisted Content Delivery: A Cooperative Approach
	%On Utilizing Meta-Surfaces in Vehicular Communications: A Multi-User Scenario For Indirect Links in Dark Zones
	Reconfigurable Intelligent Surface Enabled Vehicular Communication: Joint User Scheduling and Passive Beamforming
	
	%Cache-Enabled UAV Trajectory Planning for Maximizing Energy Efficiency using Proximal Policy  Optimization}
	%
	%A Caching Model of Internet-less UAV-Assisted Content Delivery in ITS
	}

\DeclarePairedDelimiter\abs{\lvert}{\rvert}%
\DeclarePairedDelimiter\norm{\lVert}{\rVert}%
	\newcommand{\INDSTATE}[1][1]{\State\hspace{#1\algorithmicindent}}

	\makeatletter
\newcommand{\multiline}[1]{%
  \begin{tabularx}{\dimexpr\linewidth-\ALG@thistlm}[t]{@{}X@{}}
    #1
  \end{tabularx}
}

\newcommand{\thickhat}[1]{\mathbf{\hat{\text{$#1$}}}}
\newcommand{\thickbar}[1]{\mathbf{\bar{\text{$#1$}}}}
\newcommand{\thicktilde}[1]{\mathbf{\tilde{\text{$#1$}}}}

\makeatother	
	
	\author{\IEEEauthorblockN{Ahmed Al-Hilo\IEEEauthorrefmark{1}, Moataz Samir\IEEEauthorrefmark{1}, Mohamed Elhattab\IEEEauthorrefmark{1}, Chadi Assi\IEEEauthorrefmark{1}, and Sanaa Sharafeddine\IEEEauthorrefmark{2}}\\
    \IEEEauthorblockA{\IEEEauthorrefmark{1}Concordia University,}
    \IEEEauthorblockA{\IEEEauthorrefmark{2}Lebanese American University}
	}

	\maketitle

\begin{abstract}

%In this paper, we explore meta-surface principle employment to assist vehicular communication and offer appreciable quality of service in multi-user scenarios. Specifically, we show how meta-surfaces, which are also known as reconfigurable intelligent surfaces (RIS) or intelligent reflecting surfaces (IRS), can be implemented and placed along with the logistical challenges faced in this industry.
%Owing to its ability to control wireless environments and passively improve spectral efficiency
Given its ability to control and manipulate wireless environments, reconfigurable intelligent surface (RIS), also known as intelligent reflecting surface (IRS), has emerged as a key enabler technology for the six-generation (6G) cellular networks. In the meantime, vehicular environment radio propagation is negatively influenced by a large set of objects that cause transmission distortion such as high buildings. Therefore, this work is devoted to explore the area of RIS technology integration with vehicular communications while considering the dynamic nature of such communication environment. Specifically, we provide a system model where RoadSide Unit (RSU) leverages RIS to provide indirect wireless transmissions to disconnected areas, known as dark zones. Dark zones are spots within RSU coverage where the communication links are blocked due to the existence of blockages. In details, a discrete RIS is utilized to provide communication links between the RSU and the vehicles passing through out-of-service zones. Therefore, the joint problem of RSU resource scheduling and RIS passive beamforming or phase-shift matrix is formulated as an optimization problem with the objective of maximizing the minimum average bit rate. The formulated problem is mixed integer non-convex program which is difficult to be solved and does not account for the uncertain dynamic environment in vehicular networks. Thereby, we resort to alternative methods based on Deep Reinforcement Learning to determine RSU wireless scheduling and Block Coordinate Descent (BCD) to solve for the phase-shift matrix, \textit{i.e.,} passive beamforming, of the RIS. The Markov Decision Process (MDP) is defined and the complexity of the solution approach is discussed. Our numerical results demonstrate the superiority of our  proposed approach over baseline techniques.
\end{abstract}


\section{Introduction}
Vehicular communications are deemed as integral component for the Intelligent Transportation Systems (ITS) that allow automobiles to stay connected with their surroundings as well as remote entities. They aim to provide anytime-anywhere connectivity to enable a wide range of critical and convenient services for vehicles. Indeed, the emergence of sophisticated vehicular communication technologies, \textit{i.e.}, cellular vehicle-to-everything (C-V2X) or Internet of Connected Vehicle (IoCV) \cite{chetlur2019coverage, al2020uav}, are forecast to effectively contribute in paving the way to support a plethora of essential applications including safety and non-safety related such as HD Maps, autonomous driving, 4K Video streaming, virtual and augmented reality, to name a few. Such applications urge for strict requirements such as higher throughput, lower latency, and massive connectivity. However, in the context of vehicular environment, it is strenuous to offer seamless communication experiences with ubiquitous connectivity and to enhance the quality of service.

%due to the presence of large number of various obstacles besides the highly dynamic nature of the environment.

%This paradigm enables vehicles to connect with each others in addition to external entities, vehicle to everything (V2X), i.e., Road-Side Units (RSU). Meanwhile, quality of service in next generation mobile networks, dubbed as 5G and beyond, plays a pivotal role in the future of vertical industry. It is expected that 5G will support higher throughput, lower latency, and massive connectivity. This significant advent will offer a platform that hosts essential safety and non-safety applications such as HD Maps, autonomous driving, 4K Video streaming, IoTs, VR/AR, and many others. 

Technically speaking, in many areas where large objects, \textit{i.e.}, high-rise buildings or trucks, appear, it is very probable that wireless links between terrestrial infrastructures and vehicles face frequent disturbances. Hence, the service quality falls below the desirable levels and sometimes for extended periods of time. Moreover, certain regions, where obstacles severely block Line of Sight (LoS), are permanently out of coverage which, here, are dubbed as dark zones. %In the meantime, there exist a number of services that require continuous connection in order to maintain desirable levels of Quality of Experience. For example, video streaming which includes live streaming and Video of Demand, urge for uninterrupted and seamless watchability. 
Expanding wireless coverage to unserved areas translates to dramatic raise in costs. Meanwhile, recently, reconfigurable intelligent surfaces (RIS) have been recognized as a key promising technology for achieving cost- and energy-efficient communications via smartly reshaping the wireless propagation environment \cite{wu2019towards}. RIS is composed of a number of passive low-cost elements, each of which has the ability to independently tune the phase-shift of the incident radio waves. By adequately configuring the phase-shifts with the assistance of the RIS controller, the reflected signals can be constructively added \cite{Elhattab2020Reconfigurable}. Thus, the received signal strength can be improved at the point of interest.

%Yet, a cheap, and effective solution is offered by intelligent reflecting surfaces (RIS). RISs have emerged recently to improve connectivity of wireless networks by extending communication coverage to dark zones. They act as passive relays where radio signals are reflected by adjusting RIS angle coefficients. To do so, RISs comprise a number of low-cost and passive antennas where their phase-shifts can be programmably controlled. 

 Consequently, by leveraging RIS, an indirect LoS  wireless communication link can be provided for vehicles travelling in a dark zone; \textit{i.e}. a road where large buildings block the LoS of the Road Side Unit (RSU) \cite{di2020hybrid}.  It is assumed that those vehicles are requesting service and the RSU is interested in maximizing the quality-of-service (QoS) for the passing by vehicles. To this end, the RSU operator may need to jointly optimize the RSU resource scheduling and RIS element coefficients (passive beamforming) such that the minimum average bit rates of vehicles is maximized. In addition, despite that a few works have addressed RIS phase-shift configuration in vehicular networks, only non-practical phase-shift RIS case is considered where RIS elements can have continuous element tuning. However, due to limited hardware, phase-shift elements of RIS can only have limited number of values  \cite{zhang2020reconfigurable, wu2019beamforming}.
%which are indicated by $2^b$ where $b$ represents the number of bits to control each element
%. \textcolor{red}{This direction of work is usually denoted by discrete or practical phase-shift RIS which is known to be hard to solve due to the constraint phase-shifts
%In this correspondence, we propose practical RIS-based system where a discrete RIS reflects the Road Side Unit (RSU) signals to support indirect wireless transmissions for vehicles travelling in a dark zone; i.e. a street where a large building blocks the LoS of the RSU \cite{di2020hybrid}. 

Leveraging RIS in highly dynamic environments similar to vehicular communications implies a multitude of challenges. First, vehicles constantly change their position, hence, the distance between the RIS and vehicles is varying over time and that would highly affect the channel quality between them. %For instance, when a vehicle arrives to the dark zone, it is near the RIS, meaning, as the distance is short, the established link is better. However, as the vehicle moves far into the dark zone, the gap between the RIS and vehicle becomes larger. Although, it might seem to be a good practice to serve vehicles while they are closer to the RIS, it is not always possible as the RSU has to consider all the vehicles residing within the area of interest. 
Second, the RSU has limited resources in terms of the number of available wireless channels. Thereby, the RSU needs to optimize the radio scheduling while considering the mobility of vehicles which makes the problem more challenging especially when accounting for multi-user scenarios \cite{alwazani2020intelligent}. Third, vehicles move at different and varying speeds, that is, vehicles have various residence times. Considering the same service amounts for all the vehicles passing by the dark zones will deteriorate the performance of low speed vehicles. Subsequently, maximizing the minimum average bit rates provided to navigating vehicles regardless of their sojourn times should be considered. Fourth, the arrival times and speed of upcoming vehicles are not available, in practice, upfront to the RSU operator which makes the problem further more intricate. Finally and most importantly, discrete RIS phase-shift matrix configuration is a well-known problem which is generally hard to be solved especially in a context where the phase-shift matrix of the RIS together with wireless scheduling are jointly optimized.

To the best of our knowledge, this work is the first to consider practical/discrete RIS in vehicular networks where the mobility of vehicles together with the environment uncertainties are addressed. To this end, to tackle the aforementioned challenges, an intelligent solution approach is proposed, namely Deep Reinforcement Learning, along with effective optimization technique based on block coordinate descent (BCD). The contributions of this work can be summarized as follows:
\begin{itemize}
    \item A system model is presented that leverages discrete phase-shift RIS technology to extend and enhance RSU communication. Precisely, a RSU provides service for vehicles passing through a blocked zone indirectly by employing a RIS where the mobility of vehicles and future arrivals are considered. 
    
    %We also explore the practicality of utilizing RIS in this environment taking into account vehicle dynamics and showing that RIS can improve wireless experiences in non-statistic environments.

    \item We investigate the joint vehicle scheduling and passive beamforming in RIS-empowered vehicular communication. This framework is formulated as an optimization problem with the goal of maximizing the minimum achievable bit rate for the vehicles passing through the dark zone. However, the  formulated  problem  ends  up  to mixed integer non-convex problem, which is known to be difficult to
    solve.
    %The formulated problem cannot be solved directly via traditional optimization techniques
    \item In order to tackle this challenge, we decouple the formulated problem into two sub-problems; wireless scheduling sub-problem and phase-shift matrix optimization sub-problem. Then, we resort to solve the first sub-problem via Deep Reinforcement Learning (DRL). To do so, the Markov Decision Process (MDP) is defined to be solved via DRL algorithm. Further, we propose BCD to solve the second sub-problem. We also demonstrate the robustness of our BCD algorithm. And, the computational complexity of the proposed algorithms are analyzed.
    
    \item Two case studies are carried out. The first one is to investigate how recent vehicular technologies can enable RIS integration with vehicular communications through obtaining precise vehicle positioning. Also, another study explores the area of RIS placement to optimize the overall network performance.
    
    \item Several extensive simulation based experiments are conducted using Simulation of Urban MObility (SUMO) to validate the effectiveness of our solution method and to compare with counterpart methods.
\end{itemize}

The remaining of this paper is organized as follows. Section \ref{sec:related-work} presents the major contributions that have been done in similar contexts. In Section \ref{sec:system-model}, we discusses our system model. Section \ref{sec:mathematical-formulation} formulates the problem mathematically along with the objective function. Next, Section \ref{sec:solution-approach} explains our solution approach in details. In section \ref{section:case-study}, two case studies are discussed regarding RIS placement and vehicle positioning. Then, section \ref{sec:numerical-result} shows our numerical results. Finally, we sum up the paper in Section \ref{sec:conclusion}.

\textit{Notations:} Vectors are denoted by bold-face italic letters. $\diag(x)$ denotes a diagonal matrix whose diagonal element is the corresponding element in $x$. $\mathbb{C}^{M \times N}$ denotes a complex matrix of $M \times N$. For any matrix $M$, $M^H$ and $M^T$ denote its conjugate transpose and transpose, respectively. $Pr(A \mid B)$ denotes the probability of event $A$ given event $B$.

\section{Related Work}
\label{sec:related-work}
Lately, high research efforts have been devoted towards investigating the introduction of the RIS to vehicular networks. In \cite{chen2020resource}, the authors studied resource allocation of RIS-aided vehicular communications where they aim to maximize vehicle to base station link quality while guaranteeing vehicle to vehicle communications. The authors of \cite{wang2020outage} provided analysis for outage probability in RIS-enabled vehicular networks. This paper derives an expression of outage probability showing that RIS can reduce the outage probability for vehicles in its vicinity. The analysis also proves that higher density roads increase outage probability since passing vehicles can block the communication links. %The survey paper of \cite{liu2020reconfigurable} suggests that RIS can act as a complement for On Board Unit to help autonomous and connected vehicles to receive real-time traffic information from base stations. Mainly, this work focuses on RIS assisting safety applications in vehicular networks where reliability, as always, is the main concern. 
In \cite{dampahalage2020intelligent}, the authors proposed RIS-aided vehicular networks while considering two scenarios to estimate the channels. The first one is by assuming fixed channel estimation within a coherence time. While the second one neglects the small scale fading based on the fact that vehicular positions can be realised in advance. \cite{you2020channel} considered constraint discrete phase-shift RIS with two challenges; channel estimation and passive beamforming. 

Another body of works on RIS deals with practical considerations of discrete RIS elements. In \cite{wu2019beamforming}, the authors introduced a finite number of phase-shift elements of RIS where the power is minimized while maintaining certain signal-to-interference-plus-noise ratio threshold. \cite{xu2020reconfigurable} proved how discrete phase-shift RIS is able to achieve high performance with minimum required number of  phase quantization levels. This work shows that 3 levels are enough to attain the full diversity order. In addition, the authors of \cite{di2020practical} also worked on practical RIS where multiple users are served in parallel. The objective of this work is to maximize the sum rate where the continuous digital beamforming and discrete RIS beamforming are done. \cite{tang2020mimo} proposed RIS to assist multiple-input multiple-output (MIMO) systems with 2-bit phase-shift elements. In \cite{yuan2020intelligent}, the authors proposed utilizing RIS in cognitive radio systems yielding improved spectral efficiency and energy efficiency. \cite{yan2020passive} maximized the achievable sum rate of multi-users while the RIS sends information via controlling the reflecting modulation. \cite{hu2020location} proposed a new location-based RIS where users' locations are not perfectly known. Hence, the angle between the users and RIS are estimated to configure the beams of the RIS and trasmitter. In addition, some other works also leverage RIS for security purposes, for example \cite{ai2020secure} suggested that RIS can help in alleviating security breaches related to eavesdropping. In \cite{mensi2020physical}, the authors studied the security issues related to eavesdropping attacks under different circumstances including active and passive relays (RIS).

As opposed to the previous papers, this work accounts for vehicles mobility where vehicles constantly change their position with time. Additionally, as time progresses, new vehicles arrive to the concerned area while others depart. This process of birth-and-death vehicles brings many uncertainties to the context which are hard to cope with. Thus, we aim to find a solution approach that can handle the dynamic nature of this context besides anticipating the upcoming arrivals and other hidden information about the environment. Accounting for these two objectives will help the RSU-RIS to better decide when and how to serve vehicles during their residence time. Moreover, unlike many existing works in the literature, we propose to use practical/discrete RIS.

\section{System Model}
\label{sec:system-model}
\begin{figure}
	\centerline{\includegraphics[scale=0.7]{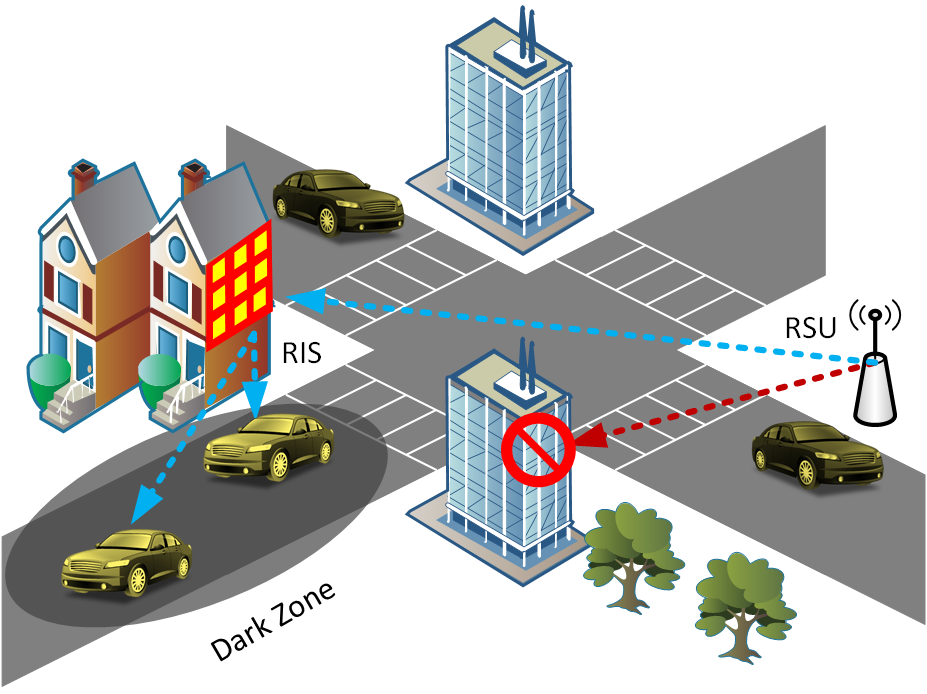}}
	\caption{System Model}
	\label{fig:system_model}
\end{figure}

We consider a particular road segment with no direct connectivity via a RSU as depicted in Fig. \ref{fig:system_model}. The line of sight (LoS) is assumed to be blocked by an obstacle, \textit{i.e.,} a high building \cite{long2020reflections}. We also consider a predefined time horizon of length $N$ which encompasses several smaller time slots, $[0,1,.., n,...,N]$. Meanwhile, we assume a flow of vehicles indexed by $v$ is navigating and requesting communication services from the RSU located at $(x_R, y_R, z_R)$ where $x_R, y_r$ are the Cartesian coordinates and $z_R$ is height of the infrastructure. The vehicles are moving at different and varying speeds, therefore, at each time slot $n \in N$, vehicle $v$ location is denoted by $(x_v^n, y_v^n, z_v)$. In order to provide uninterrupted service, the network operator leverages an RIS equipped with $M$ elements, which is situated on a building and possesses a strong LoS with both the moving vehicles passing by the dark zone and the RSU. Here, we denote the RIS location by $(x_I, y_I, z_I)$. The RSU operator aims to satisfy the vehicles by providing favourable quality of service. 

%Inspired by wireless access in vehicular environments a.k.a WAVE, 
The RSU is assumed to have a number of channels $C$ to be scheduled for the vehicles \cite{wang2011ieee}\footnote{For simplicity, we assume each vehicle can only be served via one channel.}. In case when there are several vehicles present, the RSU has to determine how to schedule its resources and tune the RIS elements. Further, due to the mobility nature of the vehicular environment the distances between the RIS and vehicles change as time progresses. Meaning, the network operator has to take into consideration that link quality degrades as vehicles moving far from the RIS.

Unlike prior work which deals with continuous phase-shift RIS, we assume a realistic scenario of a RIS where phase-shift coefficients are discrete. This scenario is more practical since it accounts for the real-world hardware limitations. However, discrete RIS is more challenging due to the additional constraint of discrete phase-shift. Moreover, the RIS consists of $M$ elements, $[1,...,m,...,M]$, each of which is controlled via $b$ bits. Hence, each one can be tuned to one of $2^b$ different angles.

\subsection{Communication Model}
In the proposed model, we consider a uniform linear array (ULA) RIS \cite{long2020reflections}. In addition, similar to the RSU, the RIS is assumed to have a certain height, $z_I$.  The communication links between RSU and RIS and that between RIS and vehicle $v$ are assumed to have a dominant line-of-sight (LoS). Thus, these communication links experience small-scale fading which are modeled as Rician fading with pure LoS components \cite{abdullah2020hybrid, samir2020optimizing}. Consequently, the channel gain between the RSU and RIS, $\boldsymbol{h}_{I,R} \in \mathbb{C}^{M \times 1}$, can be formulated as follows.

\begin{equation}
    \boldsymbol{h}_{I,R} = \underbrace{\sqrt{\rho (d_{I, R})^{-\alpha}}}_{\mathrm{path~loss}}  \underbrace{\sqrt{\dfrac{K}{1+K}} \boldsymbol{\bar{h}}_{I, R}^\mathrm{LoS}}_{\mathrm{Rician~fading}},
\end{equation}
where $\rho$ is the average path loss power gain at reference distance $d_0=1$m. Also, $K$ is the Rician factor and $\boldsymbol{\bar{h}}_{I, R}^\mathrm{LoS}$ is the deterministic LoS component which can be defined as follows 
\begin{equation}
\small
\label{eq:channel-gain-I-R}
\boldsymbol{\bar{h}}_{I, R}^{\mathrm{LoS}} = \underbrace{\Bigg[1, e^{-j\frac{2\pi}{\lambda} d \phi_{I, R}},..., e^{-j\frac{2\pi}{\lambda} (M-1) d \phi_{I, R}}\Bigg]^{T}}_{\mathrm{array~response}}, \forall n \in N,
\end{equation}
where $d_{I, R}$ is the Euclidean distance between the RIS and RSU and can be computed from $\sqrt{(x_I - x_R)^2 + (y_I - y_R)^2 + (z_I - z_R)^2}$. $\phi_{I, R}=\dfrac{x_I - x_R}{d_{I, R}}$ is cosine of the angle of arrival of signal from RSU to RIS. $d$ is the separation between RIS elements and $\lambda$ is the carrier wavelength.  

Similarly, we can compute the channel gain between the RIS and vehicles which is denoted by $\boldsymbol{h}_{I,v}^n \in \mathbb{C}^{M \times 1}$ as in Eq (\ref{eq:channel-gain-I-v}). 
\begin{equation}
\label{eq:channel-gain-I-v}
    \boldsymbol{h}_{I, v}^n = \underbrace{\sqrt{\rho (d_{I, v}^n)^{-\alpha}}}_{\mathrm{path~loss}}  \underbrace{\sqrt{\dfrac{K}{1+K}} \bar{h}_{I, v}^{n~\mathrm{LoS}}}_{\mathrm{Rician~fading}}, \forall v, n \in N, 
\end{equation}
\begin{equation}
\small
\begin{split}
  \boldsymbol{\bar{h}}_{I, v}^{n~\mathrm{LoS}} = \underbrace{\Bigg[1, e^{-j\frac{2\pi}{\lambda} d \phi_{I, v}^n},..., e^{-j\frac{2\pi}{\lambda} (M-1) d \phi_{I, v}^n}\Bigg]^{T}}_{\mathrm{array~response}}, \\ \forall v, n \in N,  
\end{split}
\end{equation}
where $d_{I, v}^n$ is the euclidean distance between the RIS and vehicle $v$ at time slot $n$ and $\phi_{I, v}^n=\dfrac{x_I - x_v^n}{d_{I, v}^n}$. Finally, we assume the channel is completely blocked between the RSU and vehicles in that zone similar to \cite{long2020reflections}\footnote{In this work, we assume that the channel gain between RIS and vehicles is fixed within one time slot.}.

%Now, the RIS contains a number of elements $M$ that can be configured at each time slot $n \in N$. 
Denote the phase-shift matrix of the RIS in the $n$th time slot as $\boldsymbol{\theta}^n = \diag\{e^{j\theta_1^n},..., e^{j\theta_M^n}\}$, where $\theta_m^n$ is the phase-shift of the $m$th reflecting element $m = 1, 2, · · · , M$. Due to the hardware limitations, the phase-shift can only be selected from a finite set of discrete values. Specifically, the set of discrete values for each reflecting RIS element can be given as $\theta_m^n \in \Omega = \{0, \frac{2\pi}{Q}, \dots, \frac{2\pi(Q - 1)}{Q}\}$, where $Q = 2^b$ and $b$ is the number of bits that control the number of available phase-shifts for the RIS elements. Hence, the signal to noise ratio (SNR) is:
\begin{equation}
\label{eq:snr}
\lambda^n_v = \dfrac{P\abs{\boldsymbol{h}_{I, R}^H  \boldsymbol{\theta}^n \boldsymbol{h}_{I, v}^n}^2}{\sigma^2}, \forall v, n \in N,
\end{equation}
where $P$ is the transmission power of the RSU and $\sigma^2$ is the thermal noise power.

Then, we can compute $\bold{h}_{I, R}^H  \boldsymbol{\theta}^n \boldsymbol{h}_{I, v}^n$ based on Eq (\ref{eq:channel-gain-I-R}) and Eq (\ref{eq:channel-gain-I-v}).
\begin{equation}
\label{eq:irs-channel-gain-expression}
\begin{split}
\boldsymbol{h}_{I, R}^H \boldsymbol{\theta}^n \bold{h}_{I, v}^n = \dfrac{\rho \dfrac{K}{K+1}}{\sqrt{(d_{I,i}^n)^{\alpha}} \sqrt{(d_{I,R})^{\alpha}}} \times ~~~~~~~~~~~~~~~~~~~~\\ \sum_{m=1}^M e^{j(\theta_m^n + \frac{2\pi}{\lambda} (m-1) d \phi^n_{I,v} - \frac{2\pi}{\lambda} (m-1) d \phi_{I,R})}, \forall v, n \in N.
\end{split}
\end{equation}

Now, instantaneous bit rate given to each vehicle is calculated as.
\begin{equation}
l_v^n = j_v^n \log_2(1 + \lambda^n_v), \forall v, n,
\end{equation}
where $j_v^n \in [0,1]$ is a decision variable to schedule the resources of RSU to vehicle $v$ at time slot $n$. Hence, $j_v^n=1$ means vehicle $v$ is served at time slot $n$ and 0 otherwise. Now, the average bit rate each vehicle receives throughout its sojourn time can be computed by the following.
\begin{equation}
z_v = \dfrac{1}{H_v}\sum_{n=1}^N l_v^n, \forall v,
\end{equation}
where $H_v$ is the residence time of vehicle $v$ in the dark zone. Next, we formally define our problem as:

\textbf{Definition 1} \textit{Assume a flow of vehicles travelling through a dark zone. The vehicles are demanding connection to a remote RSU. Meanwhile, a RIS is deployed at specific point, i.e., on a building, where it possesses a strong LoS with the RSU and the dark zone. The RIS has a certain number of elements where the operator can tune their coefficients to provide service for vehicles in order to enhance channel gains and improve bit rates. During a certain time horizon (encompassing multiple time slots), what is the best RSU wireless scheduling and phase-shift configuration for the RIS elements such that the minimum average bit rate provided to vehicles is maximized.}

\section{Mathematical Formulation}
\label{sec:mathematical-formulation}
In this section we formulate the problem of RSU wireless scheduling and RIS element tuning mathematically. Let $A_v$ and $D_v$ denote the arrival time and departure time of vehicle $v$, respectively. The notations used in this corresponding are listed in Table \ref{table:mathematical_notations}. 

\label{mathmatical_formulation}
\begin{table}[t]
	\caption{Mathematical notation}
	\begin{center}
		\begin{tabular}{|c|p{6cm}|}
		    \hline
    		\rowcolor{lightgray} \multicolumn{2}{|c|}{Parameters} \\
			\hline
			$x_I, y_I, z_I$& RIS location \\
			\hline
			$x_R, y_R, z_R$& RSU location \\
			\hline
			$x_v^n, y_v^n, z_v^n$& Vehicle $v$ location at time slot $n$ \\
			\hline
			$N$& Time horizon consists of smaller time slots. \\
			\hline
			$V^n$& Set of available vehicles during time slot $n$ \\
			\hline
			$C$& Number of RSU channels \\
			\hline
			$H_v$& Vehicle $v$ residence time \\
			\hline
			$\phi_{I,R}$& Angle of arrival at RIS from RSU \\
			\hline
			$\phi_{I,v}^n$& Angle of arrival between the RIS and vehicle $v$ at time slot $n$ \\
			\hline
			$\alpha$& Path loss exponent \\
			\hline
			$P$& Transmission power of the RSU \\
			\hline
			$\rho$ & Median of the mean path gain at reference distance = 1m \\
			\hline
			$\sigma$ & Thermal noise power \\
			\hline
			$b$ & Number of control bits for the RIS elements\\
			\hline
			$Q$ & Number of RIS phase-shift patterns \\
			\hline
    		\rowcolor{lightgray} \multicolumn{2}{|c|}{Variables} \\
			\hline
			$j_v^n$& 1: if vehicle $v$ is scheduled for service by the RSU at $n$ and 0 otherwise  \\
			\hline
			$\theta^n_m$& RIS element $m$ phase-shift angle at time slot $n$\\
			\hline
		\end{tabular}
		\label{table:mathematical_notations}
	\end{center}
\end{table}

%Let us define $S=[s_1, s_2,...s_v,...,s_V]$ as a vector contains all average bit rate provided for the vehicle. 
The optimization problem alongside the objective function can be mathematically written as follows. For the sake of clarity, let $\boldsymbol{\Theta} = \{\boldsymbol{\theta}^1, \boldsymbol{\theta}^2, \dots, \boldsymbol{\theta}^N\}$ and $J=\{j_v^n, \forall v, n \in N\}$.
\begin{maxi!}|s|[2]
    {\boldsymbol{\Theta}, J}
    {\{\min z_v\} \label{eq:objective}} 
    {\label{eq:Example1}} 
    {} 
    \addConstraint{}{\sum_{v=1}^{V^n} j_v^n \leq C, \forall n \in N, \label{eq:con1}}  
    \addConstraint{}{j_v^n \in [0,1], \forall n \in N, v, \label{eq:one-vehicle}}
    \addConstraint{}{j_v^n \leq \max(n - A_v,0), \forall n \in N, v, \label{eq:con2}}
    \addConstraint{}{j_v^n \leq \max(D_v - n,0), \forall n \in N, v, \label{eq:con3}}
    \addConstraint{}{\theta_m^n \in \Omega, \forall n \in N, m \in M. \label{eq:con4}}   
\end{maxi!}

Here, the objective function, Eq (\ref{eq:objective}), is max-min which translates to maximizing the minimum average bit rate. Constraint (\ref{eq:con1}) ensures that the number of channels scheduled to vehicles is no more than that available at the RSU. Constraint (\ref{eq:one-vehicle}) allows vehicles to be served via one channel only. Constraints (\ref{eq:con2}) and (\ref{eq:con3}) make sure that vehicle can only be served via RIS while it is within the area of the dark zone. Finally, constraint (\ref{eq:con4}) restrains the number of phase-shift values. Now, the problem is non-convex due to the discrete RIS element phase-shift optimization. Also, the phase-shift matrix is hard to be solved. For instance, if the phase-shift is tuned to optimally serve the first vehicle, the other ones might receive less quality and vise versa. Furthermore, in this problem, it is hard to eliminate the coupling relationship between phase-shift configuration and wireless scheduling. In addition, the information of vehicles such as their arrival, speed, and departure, are unknown in advance. Due to the dynamic nature of the environment, it is impractical to assume such information is given. Hence, a effective solution mechanism has not only to deal with the difficulties of such problem, but it has also to predict for the hidden parameters. 

In order to address the above challenges, we resort to Deep Reinforcement Learning (DRL) with multi-binary action space to find a policy that maximizes the minimum average bit rate for vehicles. However, if DRL is used to solve for the two decisions of resource scheduling and phase-shift matrix, the action space will be equal to all the possible combinations of wireless scheduling and phase-shift patterns for $M$ elements which is unbearably large. Such massive action space would increase the DRL agent difficulty to learn. Similar to \cite{lee2020deep, zhang2020distributional}, a more practical solution approach can be realised by delegating one decision to an optimization technique while dedicating the second one to machine learning based approach. In particular, the DRL agent first determines which vehicles are going to be served at time slot $n$. While, BCD algorithm is invoked to configure the phase-shift matrix such that the service offered to the scheduled vehicles is optimized. Next, the solution approach, in details, will be discussed.

%Therefore, we resort to DRL as it is an efficient technique that can deal with such context.

%However, there are still some limitation corresponding to DRL. A question might be raised, why do not we use DRL to solve both problems. The answer is that DRL is still limited by action space. Despite there are a number of enhancement done to improve its ability to deal with large action spaces, DRL can only handle a limited number of actions. Additionally, as the action space grows up, the convergence of DRL becomes slower. On the other hand, RIS is known to have large number of elements in order to attain good performance, hence, DRL needs to cope with relatively large number of actions if it is chosen to decide for the RIS phase-shift coefficients. Therefore, a more practical solution approach can be realised via off-loading the task of phase-shift configuration to an alternative method, here BCD, while dedicating DRL to solve for RSU resource scheduling. A practical study is done by \cite{lee2020deep} demonstrates that DRL can only handle small number of RIS elements and with somehow poor convergence. In the next section we will explain our solution approach in details

\section{Solution Approach}
\label{sec:solution-approach}
The solution approach for joint resource scheduling and passive beamforming is presented in this section. First, we decompose the aforementioned problem into two sub-problems, the first sub-problem is due to the resource scheduling and second one corresponds to the phase-shift matrix of the RIS. The information and mobility of the upcoming vehicles are unknown in advance. That is, solving the first sub-problem is quite challenging. Hence, we resort to DRL to observe the environment and tackle multi-user RSU scheduling. Next, the RIS elements are tuned based on Block Coordinate Descent (BCD) \cite{bai2020latency, he2020reconfigurable, abeywickrama2020intelligent}. The details of our solution methodology are laid out in the next sections.

\subsection{DRL for Wireless Scheduling}
For the DRL, we need to define a Markov Decision Process tuple $\langle\boldsymbol{S}, \boldsymbol{A}, \gamma , R, G\rangle$ that represents the environment. First, let us define four new notations; $f^n, \forall n \in N$ which denotes the current minimum average bit rate until time slot $n$, $k_v^n, \forall n \in N, v$ denotes the speed of vehicle $v$ at time slot $n$, $z_v^n \forall n \in N, v$ is the current average bit rate of vehicle $v$ until time slot $n$, $\eta$ is the largest number of vehicles existing simultaneously, and $U$ is the number of possible actions. Now, the MDP is defined as follows:
\begin{itemize}
    \item $\boldsymbol{S}$ is the state space where its size is large as it contains unbounded parameters of real numbers. The system state $s^n \in \boldsymbol{S}$ is a vector that indicates current minimum service provided up to $n-1$, the speeds of the existing vehicles ($k_v^n, \forall v$) in that time slots, their cumulative average bit rates up to $n$ ($z_v^n, \forall v$), and their locations ($x_v^n, \forall v$). The state $s$ can be expressed as:
    \begin{equation}
        \{f^n, \underbrace{k_1^n, z_1^n, x_1^n}_{v=1}, \underbrace{k_2^n, z_2^n, x_2^n}_{v=2}, ... , \underbrace{k_{\eta}^n, z_{\eta}^n, x_{\eta}^n}_{v=\eta}\}, \forall n.
    \end{equation}

    %Now, why the $f^n$ is important in the state vector. In fact, this value will tell the agent what is the worst service provided in terms of minimum average bit rate. So, the agent should try not to reduce this value as per Eq \eqref{eq:objective}. 
    
    \item $\boldsymbol{A}$ is the action space where the action taken for each time slot $n$ is $a^n \in \boldsymbol{A}$. $a^n$ is a binary vector of size $\eta$. Also, the sum of vector $a^n$ should be equal to $C$ (to enforce constraint \eqref{eq:con1}). For example, if $a^n[0]=1$, then the first existing vehicle is being served at time slot $n$ and so forth. The number of actions can be computed by $U = \eta!/(C! (\eta-C)!)$. The possible combinations of action vectors is similar to the example below. 
    \begin{equation}
        \underbrace{\{\underbrace{0}_1,0,1,...,\underbrace{1}_{\eta}\}}_{1},\underbrace{\{0,1,1,...,1\}}_{2},...,\underbrace{\{1,1,0,...,0\}}_{U}
    \end{equation}
    
    \item $R$ is the discounted cumulative reward produced after executing every step $n \in N$ and it is defined as follows:
    \begin{equation}
     R = \sum_{n=1}^{N} \gamma^{n-1} r^n, 
    \label{eq:cumulative_reward}
    \end{equation}
    
    Here, the step reward $r^n$ is computed as follows. During the beginning of the operational phase, $r^n=0$ until the first vehicle departs. For the first vehicle, the step reward is equal to its average bit rate. Henceforth, whenever any vehicle leaves the dark zone, the step reward is given as a penalty if and only if that vehicle has received less average bit rate than the other vehicles which left previously. It is worth noting that since the agent seeks to maximize the minimum average bit rate, it does not count reward if a vehicle received higher bit rate than others.
    
    \item $G$ denotes state transition probabilities which is the probability of being in state $s'$ after applying action $a^n$ in state $s^n$. The probability of transition from one state to another depends only on the current state as in Eq (\ref{eq:mdp_proof}).
    \begin{align}
    \small
        \begin{split}\label{eq:mdp_proof}
            Pr(s^{n+1} = s' \mid s^n, a^n) {}& = Pr(f^{n+1} \mid f^{n}, a^n)\\
                 %& \times  %Pr(V^{n+1} \mid V^{n}) \\
                 & \times  Pr(k_v^{n+1} \mid k_v^n) \\
                 & \times  Pr(x_v^{n+1} \mid x_v^n, k_v^n) \\
                 & \times  Pr(z_v^{n+1} \mid z_v^{n}, a^n). \\
        \end{split}
    \end{align}

    That is, the probability of having minimum average bit rate of $f^{n+1}$ depends on the current value $f^n$ and the action taken at that time slot $a^n$. %The probability of having a certain set of vehicles, $V^{n+1}$, depends only on the current set of vehicles ($V^{n}$). 
    The probability of vehicle having certain speed, $k_v^{n+1}$, depends on its current speed. Moreover, the probability of vehicle $v$ being in next location, $x_v^{n+1}$ depends on its current location ($x_v^n$) and its speed ($k_v^n$). Finally, the probability of vehicle having certain average bit rate ($z_v^{n+1}$) depends only on the current one ($z_v^{n}$) and the action taken at that time slot.
\end{itemize}

\textbf{Remark} \textit{selecting an action is a non-trivial task for the problem explained above. Actually, since our objective is to maximize the minimum average bit rate for all vehicles, it is not easy to decide which vehicles to serve at each time slot. For instance, a vehicle has just entered may have plenty of time to be served later while a vehicle near the end of the road segment may have no much time to receive service. In contrary, a vehicle located at the end is way far than those vehicles near the RIS. Hence, the latter can receive much higher bit rate if selected to be served. Moreover, if the RSU postpones the service for one vehicle, other vehicles may arrive, therefore, that vehicle will have less chances to be served later. In addition, in our work, we consider multi-user communication where more than one vehicle might be scheduled by the RSU simultaneously which makes the action space more complicated. Hence, the agent needs to interact with the environment and try different actions and scheduling policies in order to figure out which one attains the best cumulative reward.}

For DRL, we exploit PPO to develop our agent as laid out in Algorithm \ref{algorithm:ppo_algorithm}. First, the agent initializes random sampling policy and value function for the neural networks as in lines 3 and 4. Then, in each epoch, the agent observes the environment which consists of the set of vehicles and their information, minimum average bit rate achieved up to $n$. Then at each time slot $n$, the agent selects an action which is a binary vector that determine which set of vehicles will be served via the RSU. Based on that action, the BCD algorithm is then invoked to configure the phase-shift matrix in order to maximize the channel gain. Eventually, the time step reward is worked out which has three cases. First, if no vehicle has departed yet, $r^n=0$. Second, if the very first vehicle departs, the reward is set to its average bit rate. Third, the consecutive vehicles leaving the area will be accounted as a penalty if and only if their average bit rate is less than $f^n$ when they have departed.

After gathering the set of samples and computing the rewards, the agent works out the advantage function (line 19), $\hat{A}$, which is defined as the resultant of subtracting the expected value function from the actual reward. $\hat{A}$ is the estimated advantage function or relative value of the selected action. It helps the system to understand how good it is preforming based on its normal estimate function value \cite{bohn2019deep}.

Our agent is based on proximal policy optimization which usually is implemented in Actor-Critic framework, where more objective functions are added to the surrogate objective. Based on \cite{shokry2020leveraging}, the complexity of connected network with $P$ layers is  $O(\sum_{p=0}^P n_p n_{p-1})$ where $n_p$ denotes the total number of neurons in layer $p$.

    \begin{center}
    \begin{algorithm}[t]
        \small  
    	\caption{Proposed DRL for Scheduling}
    	\label{algorithm:ppo_algorithm}
    	\begin{algorithmic}[1]
    	    \State \textbf{Inputs:} $N$, $v$, Learning Rate, $\gamma$, $\epsilon$.
    	    \State \textbf{Outputs:} RSU resource scheduling and $\boldsymbol{\theta}^n$.
    	    \State Initial policy $\pi$ with random parameter $\theta$ and threshold $\epsilon$
    	    \State Initial value function $V$ with random parameters $\phi$
    	    \For{each episode $k \in \{0, 1, 2,...\}$}
    	        \For{$n:\{0, 1, 2,..., N\}$}
    	        
    	            \State Observe state $f^n, k_v^n, z_v^n, x_v^n, \forall v \in V^n$. 
    	            \State Select action $a^n$ from $\pi_{\theta_{old}}$
    	            
    	            \State Assign channel to vehicle $v$ if it is scheduled to be served.
    	            
    	            \State Configure RIS phase-shift matrix using Algorithm \ref{algortihm:BCD}.
    	            
    	            \If{Vehicle $v$ is the first one to leave}
    	                \State Set $r^n = z_v$
    	            \ElsIf{Vehicle $v$ departed and $z_v < f^n$}
    	                \State Set $r^n = f^n - z_v$
    	            \Else
    	                \State $r^n = 0$
    	           \EndIf
    	            
    	        \EndFor
    	        \State Compute advantage estimate $\hat{A}$ for all epochs.
    	        \State Optimize surrogate loss function using Adam optimizer.
    	        \State Update policy $\pi_{\theta_{old}} \gets \pi_{\theta}$.
    	    \EndFor
    	\end{algorithmic}
    \end{algorithm}
    \end{center}

\subsection{BCD for RIS Phase-Shift Coefficients}
Block coordinate descent (BCD) has been proposed in the literature to solve for RIS phase-shift matrix \cite{abeywickrama2020intelligent, he2020reconfigurable}. In this correspondence, we aim to leverage BCD to maximize the sum of immediate sum of bit rates of all vehicles selected to be served at time slot $n$.
\begin{equation}
    \label{eq:max-sum-bcd}
    \sum_{v=1}^{V^n} j_v^n l_v^n, \forall n.
\end{equation}

To do so, Algorithm \ref{algortihm:BCD} receives the action selected by DRL in Algorithm \ref{algorithm:ppo_algorithm}, $J^n$. Once the decision is taken by the agent, the BCD is then called to optimize the phase-shift matrix in iterative way. In each iteration, a sequence of block optimization procedures are performed. In each one, all elements are fixed while one is optimized by checking all its possible values, $2^b$. The one that maximizes the objective will be selected. After that, the next element will be selected to optimize and so forth. This operation is iterated until Eq \eqref{eq:max-sum-bcd} has converged. In practice, Algorithm \ref{algorithm:ppo_algorithm} needs one iteration to surpass 95\% threshold of its maximum performance. Hence, this algorithm is pretty robust in dealing with the phase-shift coefficients.

    \begin{center}
    \begin{algorithm}[h]
        \small
    	\caption{BCD to Tune the RIS Phase-Shift Matrix}
    	\label{algortihm:BCD}
    	\begin{algorithmic}[1]
    	\State \textbf{Inputs:} $J^n$.
        \State \textbf{Outputs:} $\boldsymbol{\theta^n}$
        \While{Eq (\ref{eq:max-sum-bcd}) not converged}
            \For{$m=1,...,M$}
                \State Fix $m', \forall m' \neq m, m' \in M$
                \State Set $\theta_m^n = \underset{\Omega}{\argmax}~ $ Eq(\ref{eq:max-sum-bcd})
            \EndFor
            \State Obtain Eq (\ref{eq:max-sum-bcd})
        \EndWhile
    	\end{algorithmic}
    \end{algorithm}
    \end{center}

\begin{figure}
	\centerline{\includegraphics[scale=0.25]{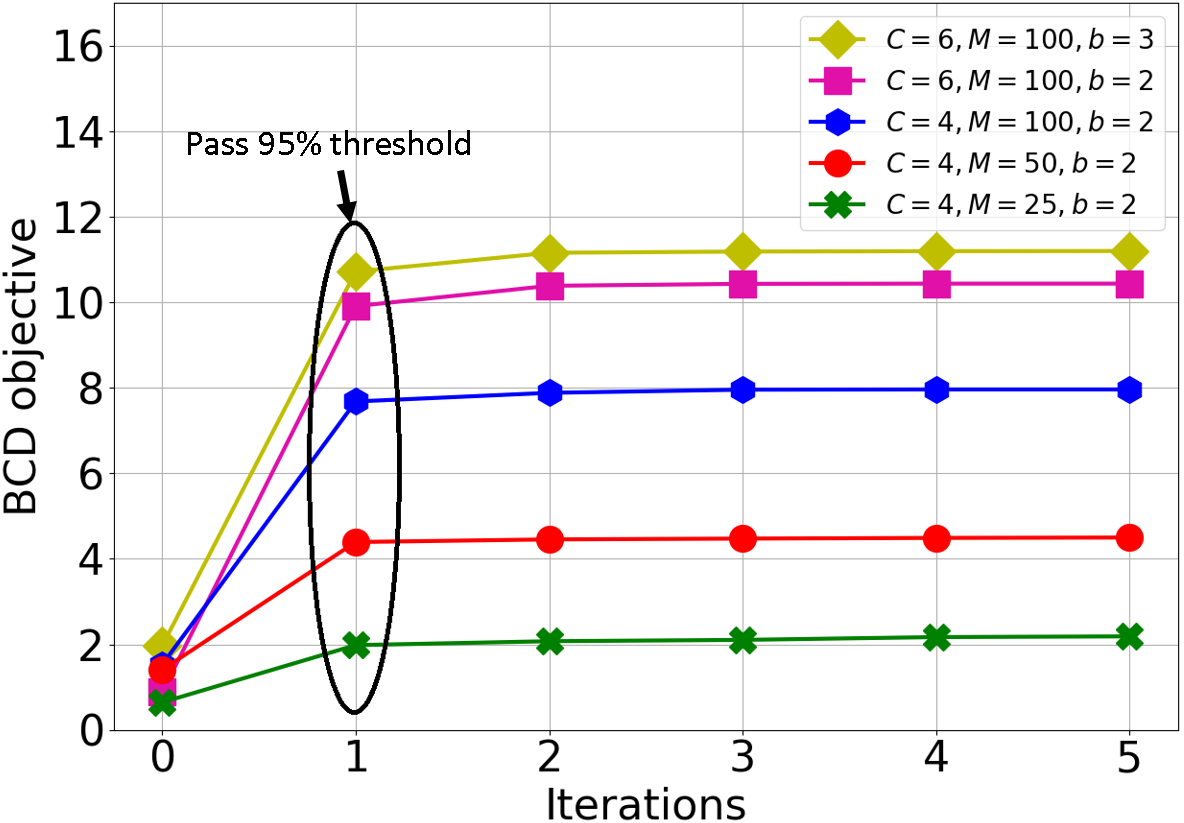}}
	\caption{BCD convergence over iterations with different RIS elements ($M$), users ($C$), and quantization levels ($b$).}
	\label{fig:bcd-iterations}
\end{figure}

Concerning the complexity of Algorithm \ref{algortihm:BCD}, it is $O(I M 2^b)$ where $I$ stands for the number of iterations until Eq \eqref{eq:max-sum-bcd} converges. In details, there are three loops in this Algorithm; first is the number of iterations, second is the number of RIS elements, and third is the number of angles available to control each element. An experiment is conducted to study the BCD performance and the results are shown in Fig. \ref{fig:bcd-iterations}. In this experiment, we vary the number of RIS elements from 25 to 100 elements. Moreover, we try different number of users ($C$) and control bits ($b$). Based on the outcomes, we can approximate the complexity of Algorithm \ref{algortihm:BCD} to $O(M 2^b)$. The complexity can further be approximated to $O(M2^3) = O(8M)$ based on the fact that a $b$ of 3 is enough \cite{zhang2020reconfigurable}\footnote{Note that, based on our experiments, we found that $b = 2$ is enough to
achieve high performance in our context as demonstrated in Section \ref{sec:numerical-result}.}.

\section{Case Study: RIS Placement and Vehicle Positioning}
\label{section:case-study}
In the literature, RIS is often proposed to tune static wireless environments such as user equipment or IoT devices. However, implementing RIS in dynamic medium is far more challenging owing to the highly sensitivity of RIS phase-shift alignment. Therefore, we carry out two studies to address the practical RIS placement and the impacts of vehicle positioning accuracy in the context of RIS-assist vehicular communications.

\subsection{RIS Placement}
In this section, we discuss the issue of placing RIS at different places. We statistically study how placing the RIS at different locations can actually improve or worsen the overall performance. Then, we will see what is the optimal location to situate the RIS. We start off by a hypothesis stating that the optimal RIS placement is the closest one to the RSU. This hypothesis is based on initial observations that indicate the shorter the distance between the RIS and RSU, the best channel gain can be achieved. In order to prove our claim, we are going to derive it mathematically and then back it up with simulation experiments.

\textbf{Theorem 1} Given $x_I < x_I^{'}, x_R < x_I, x_R < x_v, \forall v$, the inequality of $z_v(x_I) > z_v(x_I^{'}), \forall x_I < x_I^{'}$ always holds.

\begin{figure}[t]
	\centerline{\includegraphics[scale=0.25]{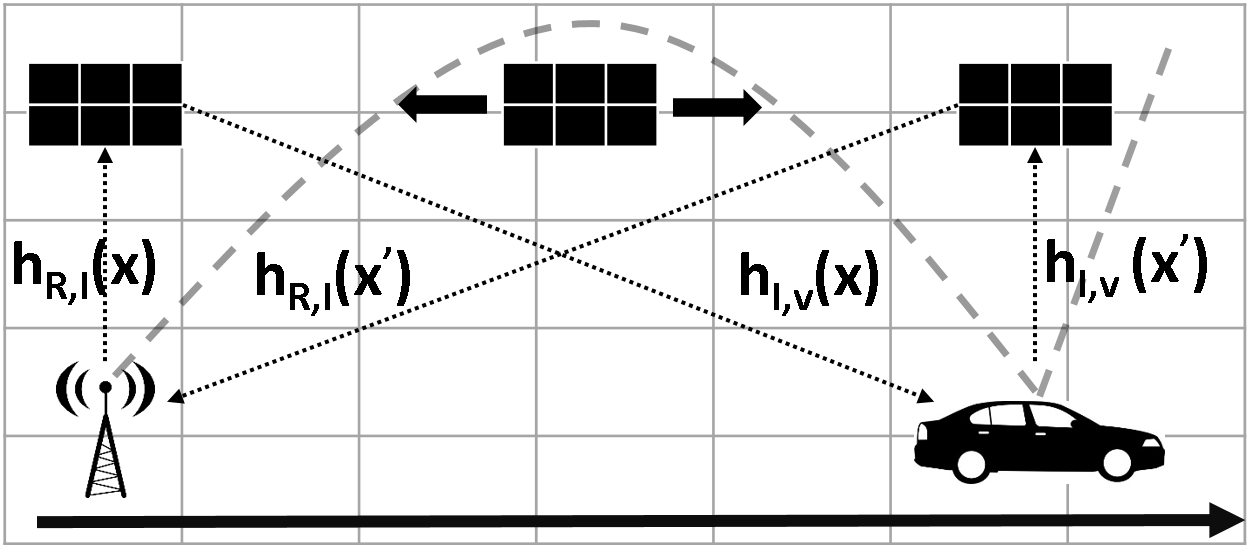}}
	\caption{RIS placement with corresponding channel gain, in the background, in 1D space.}
	\label{fig:irs-placement-illustration}
\end{figure}

\textit{Proof:} Here, for simplicity, we take a RIS and try to place it at different points to serve a single vehicle as shown in Fig. \ref{fig:irs-placement-illustration}. Hence, the RIS elements will always be tuned to maximize the channel gain for that vehicle. Fortunately, Eq \eqref{eq:irs-channel-gain-expression} can be obtained in a closed form for a single user \cite{li2020reconfigurable}.
\begin{equation}
    \theta_m^n =\frac{2\pi}{\lambda} (m-1) d \phi^n_{I,v} + \frac{2\pi}{\lambda} (m-1) d \phi_{I,R}, \forall m \in M, n.
\end{equation}

Where the phase-shifts of RIS cancel out the ones of RIS-RSU and RIS-vehicle, Eq (\ref{eq:irs-channel-gain-expression}) can be rewritten as:
\begin{equation}
\label{eq:irs-placement-derived-eq-2}
\boldsymbol{h}_{I, R}^H  \boldsymbol{\theta}^n \boldsymbol{h}_{I, v}^n = \dfrac{\rho \dfrac{K}{1+K} M}{\sqrt{(d_{I,i}^n)^{\alpha}} \sqrt{(d_{I,R})^{\alpha}}}.
\end{equation}

Since $\rho$, $M$, and $\alpha$ are constant, the only factors that remain variable are $d_{I,v}^n$ and $d_{I,R}$ which denote the distances of RSU-RIS-vehicle. We can also notice that Eq (\ref{eq:irs-placement-derived-eq-2}) is a decreasing function with respect to distances. Now, we need to prove that $z_v(x_I) > z_v(x_I^{'}), \forall x_I < x_I^{'}$. To do so, let us assume a vehicle has $H_v = N$ (we assume there is a single vehicle on the road). To this end, $z_v(x_I) > z_v(x_I^{'})$ is greater when the sum of bit rates received throughout $H_v$ is larger.
\begin{equation}
\small
\label{eq:proof-1}
\begin{split}
l_v^1(x_I) + l_v^2(x_I)+ ... + l_v^N(x_I) > l_v^1(x_I^{'}) + \\ l_v^2(x_I^{'}) + ... + l_v^n(x_I^{'}), \forall x_I < x_I^{'}.
\end{split}
\end{equation}

Next, for clarity, let $Y = \dfrac{P \rho^2 \bigg(\dfrac{K}{1+K}\bigg)^2 M^2}{\sigma^2}$ which is a invariant value. Hence, Eq \eqref{eq:proof-1} can be rewritten as:
\begin{equation}
\label{eq:simplification-1}
\small
\begin{split}
log_2(1 +  \dfrac{Y}{(d_{I,v}^1)^{\alpha} (d_{I,R})^{\alpha}}) + log_2(1 +  \dfrac{Y}{(d_{I,v}^2)^{\alpha} (d_{I,R})^{\alpha}}) + ... \\ + log_2(1 +  \dfrac{Y}{(d_{I,v}^N)^{\alpha} (d_{I,R})^{\alpha}}) > log_2(1 +  \dfrac{Y}{(d_{I,v}^{'1})^{\alpha} (d^{'}_{I,R})^{\alpha}}) + \\ log_2(1 +  \dfrac{Y}{(d_{I,v}^{'2})^{\alpha} (d^{'}_{I,R})^{\alpha}})  + ... + log_2(1 +  \dfrac{Y}{(d_{I,v}^{'N})^{\alpha} (d^{'}_{I,R})^{\alpha}}) \\ , \forall x_I < x_I^{'}.
\end{split}
\end{equation}

Now, only the distance would affect the sum of bit rates over $N$. To facilitate the expressions, let us further assume one dimensional environment. Hence, $d_{I,R} = \abs{x_I - x_R}$ and $d_{I,v}^n=\abs{x_v^n - x_I}$. We can also assume $\alpha = 1$. Therefore, as absolute values are multiplicative:
\begin{equation}
\label{eq:simplification-2}
\begin{split}
d_{I,R} \times d_{I,v}^n = \abs{x_I x_v^n - x_I^2 - x_R x_v^n + x_R x_I}.
\end{split}
\end{equation}

\begin{figure}
  \centering
  \begin{tabular}{@{}c@{}}
    \includegraphics[scale=0.26]{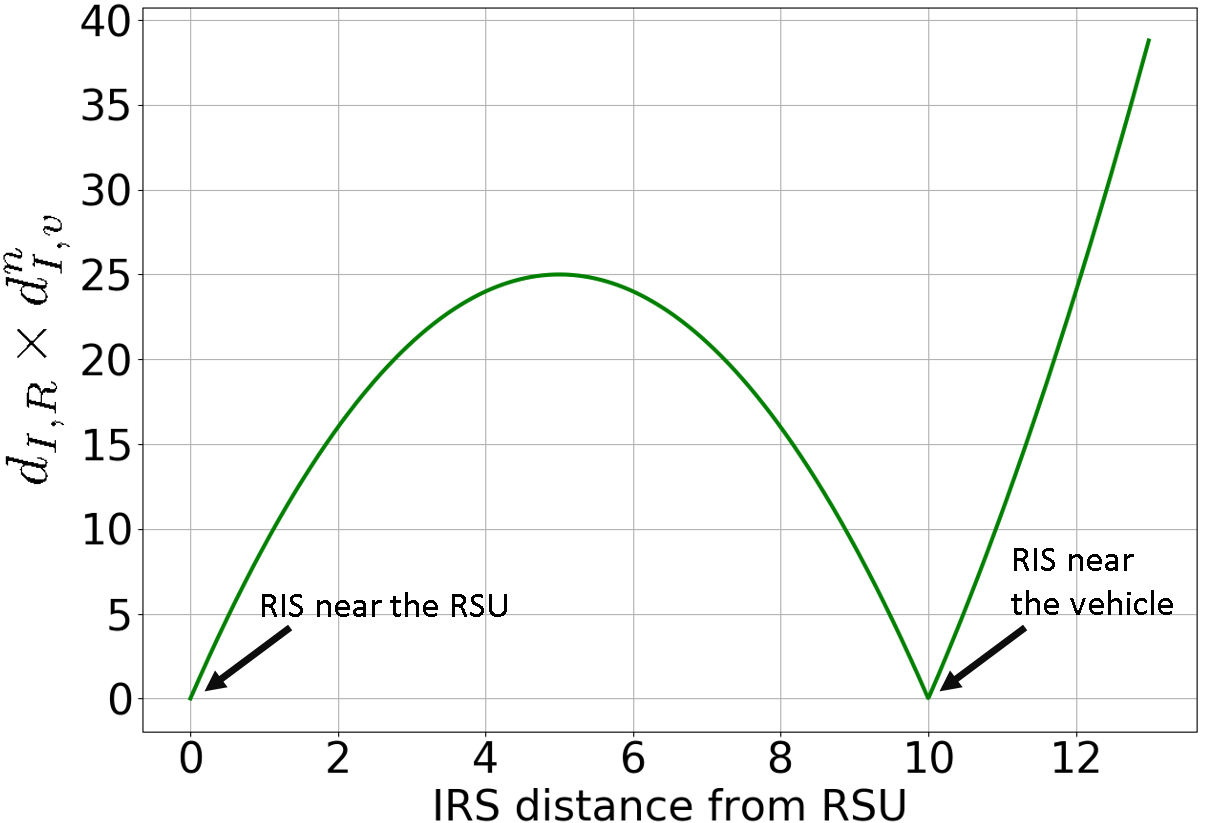} \\[\abovecaptionskip]
    \small (a) Eq (\ref{eq:simplification-1}) curve with varying $x_I$ values.
  \end{tabular}

  \vspace{\floatsep}

  \begin{tabular}{@{}c@{}}
    \includegraphics[scale=0.25]{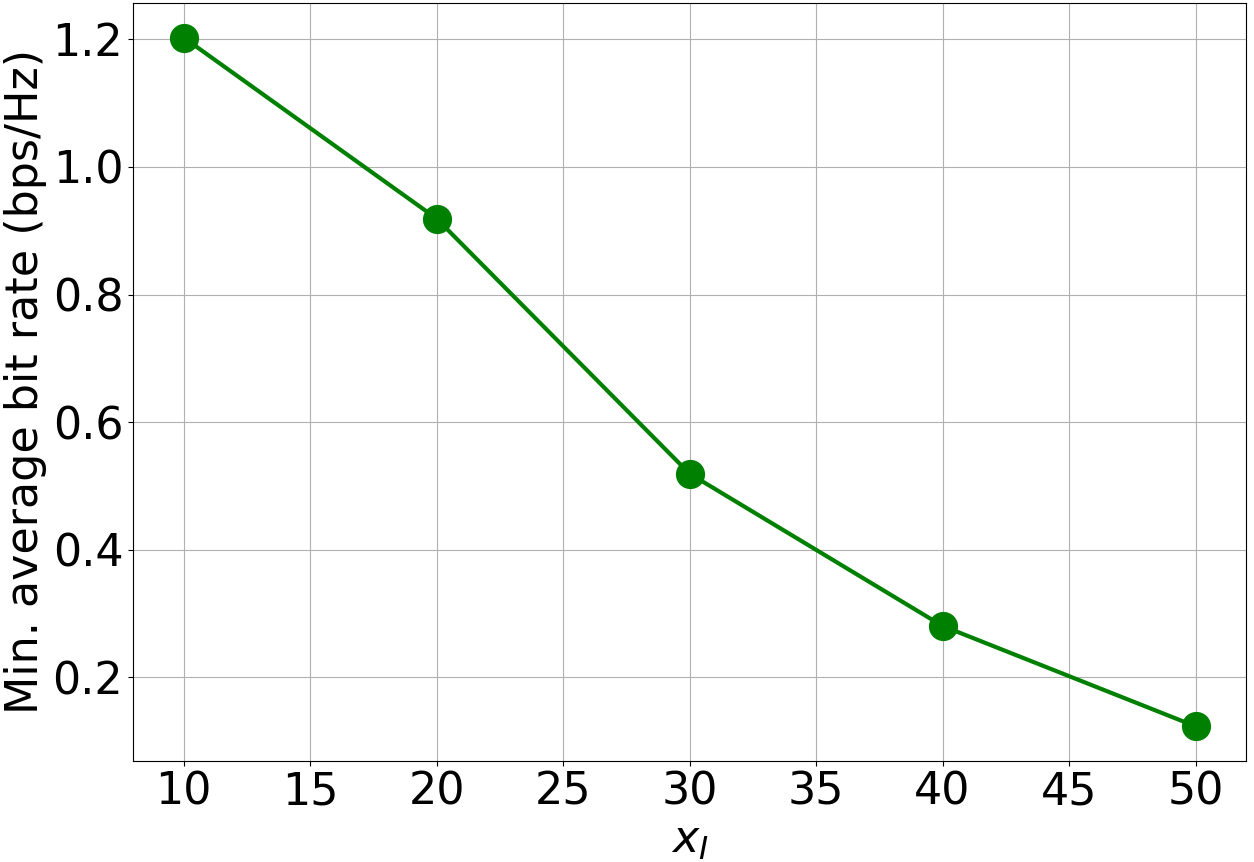} \\[\abovecaptionskip]
    \small (b) Multi-user scenario with varying $x_I$ values.
  \end{tabular}
  \caption{Simulation results for Eq (\ref{eq:simplification-1}) and RIS placement.}
  \label{fig:irs-placement-proof}
\end{figure}

In Fig \ref{fig:irs-placement-proof} (a), note that Eq (\ref{eq:simplification-2}) has its lowest value when $x_I = x_R$ or $x_I = x_v^n$. The first term can always be achieved, during the entire time horizon ($\forall n \in N$) and for all vehicles, as long as $x_I$ and $x_R$ are fixed. However, the second one, $x_I = x_v^n$, as vehicles driving, only holds true for one time slot $n$ and for a specific vehicle. Therefore, with $H_v > 1$, we can confidently say:
\begin{equation}
\small
\begin{split}
\sum_{n=1}^{N} log_2(1 + \dfrac{Y}{(d_{I,v}^n)^{\alpha} (d_{I,R})^{\alpha}}) > \sum_{n=1}^{N} log_2(1 + \dfrac{T}{(d_{I,v}^{'n})^{\alpha} (d_{I,R}^{'n})^{\alpha}}), \\ \forall x_I < x_I^{'}.
\end{split}
\end{equation}

%Therefore, each term in Eq (\ref{eq:simplification-1}) left side is greater than its counterpart in the right side.
Consequently, the right side is greater the left side which completes the proof.

In Theorem 1, we have shown that, for a single vehicle, the ideal place for the RIS is to be as closer as it can to the RSU. However, in practice, there exists several constraints that force the RIS to be distant from the RSU. For example, the LoS has to be clear between the RIS and RSU and between the RIS and the vehicles it serves. Otherwise, the wireless links would be highly disturbed and the RIS will lose its functionality. In line with our proof and discussion above, we carry out three experiments to see the achievable minimum average bit rate for multi-user scenario. We set $x_R=0$, $x_I=[10, 20, 30, 40, 50]$. Vehicles are generated by SUMO. The outcomes are displayed in Fig. \ref{fig:irs-placement-proof} (b). One can observe that with RIS closer to RSU, the performance was much higher. However, this performance started to degrade dramatically as the RIS moves away from the RSU.

\subsection{Vehicle Positioning Precision}

One of the challenging issues in vehicular communications is vehicle positioning in such highly dynamic environment. Hence, in the context of RIS-assist vehicle communication, inaccurate vehicles positioning might lead to severe consequences that negatively impact the channel gain. Thus, we attempt to understand whether it is possible to leverage the emerging technologies related to vehicle positioning such as 3D-LIDAR (Light Detection and Ranging), Global Positioning System, etc., for the benefits of accurately estimating real vehicle positions to enable RIS-aided vehicular communications.

First, based on the channel gain equations laid out in Eq \eqref{eq:irs-channel-gain-expression}, the impact of inaccurate vehicle positioning on the RIS-based system performance will be examined. Let $\Delta$ be the error in vehicles positioning, then, the estimated position is $x_v^n \pm \Delta$. It is worth noting that in this context, the small error ($\Delta$) in vehicle positioning will not have significant impact on the cascaded channel (RSU-RIS-vehicle) path loss. Thus, we only study the phase-shift angle deviation (difference between accurate and estimated angles of arrival). Next, we highlight the components in Eq \eqref{eq:irs-channel-gain-expression} that are affected by inaccurate vehicles positioning. The next equation describes the phase-shift multiplication of the two angles of arrival, $\phi_{R,I}$ and $\phi_{I,v}^n$, with the RIS elements at the real position.
\begin{equation}
\label{eq:vehicle_positioning_1}
\small
\begin{split}
    \theta_m^n + \frac{2\pi}{\lambda} (m-1) d \dfrac{x_I - x_v^n}{\sqrt{(x_I - x_v^n)^2 + Y}} - \frac{2\pi}{\lambda} (m-1) d \phi_{I,R}, \\ \forall m \in M, n \in N.
\end{split}
\end{equation}

Next, we formulate the same equation, yet, at the estimated position.
\begin{equation}
\label{eq:vehicle_positioning_2}
\small
\begin{split}
    \theta_m^n + \frac{2\pi}{\lambda} (m-1) d \dfrac{x_I - x_v^n \pm \Delta}{\sqrt{(x_I - x_v^n \pm \Delta)^2 + Y}} - \\ \frac{2\pi}{\lambda} (m-1) d \phi_{I,R}, \forall m \in M, n \in N.
\end{split}
\end{equation}

After subtracting the two equations, real and estimated one (Eq \eqref{eq:vehicle_positioning_1} - Eq \eqref{eq:vehicle_positioning_2}), we end up with the following.
\begin{equation}
\label{eq:vehicle-position}
\small
\begin{split}
\frac{2\pi}{\lambda} (m-1) d (\dfrac{x_I - x_v^n}{\sqrt{(x_I - x_v^n)^2 + Y}} - \dfrac{x_I - x_v^n \pm \Delta}{\sqrt{(x_I - x_v^n \pm \Delta)^2 + Y}}) \\, \forall m \in M, n \in N.
\end{split}
\end{equation}

Eq (\ref{eq:vehicle-position}) clearly states that the cosine angle of arrival between RIS and vehicles is affected by the error in vehicle positioning. It is also noted that this impact occurs for all the RIS elements. How much this deviation from the real position affects the performance is what we answer next. 

\begin{figure*}[htbp]
    \centering
    \subfloat[]{\includegraphics[scale=0.19]{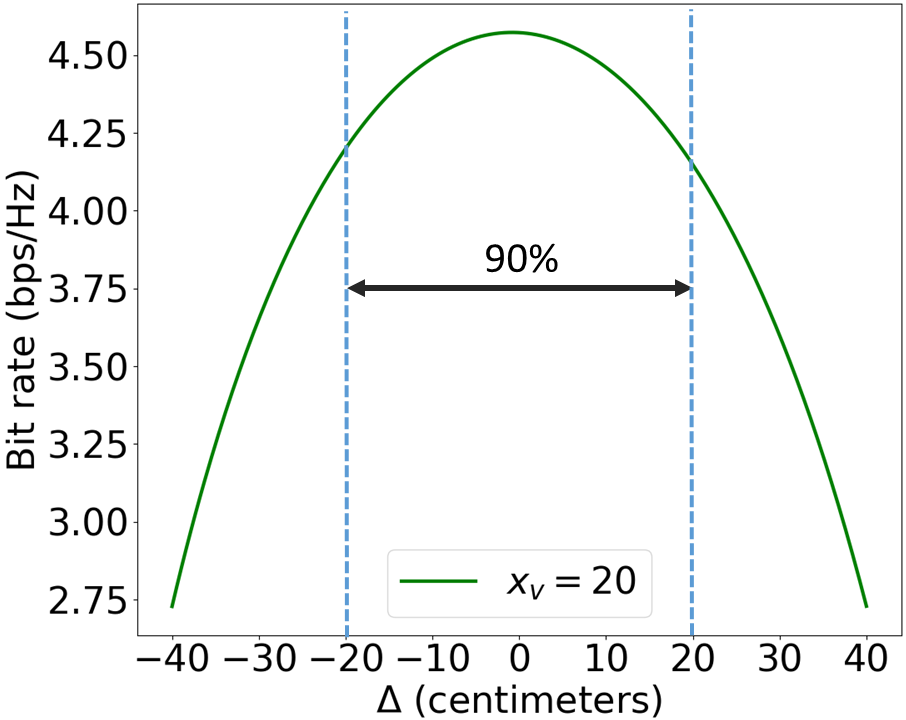}} 
    \subfloat[]{\includegraphics[scale=0.19]{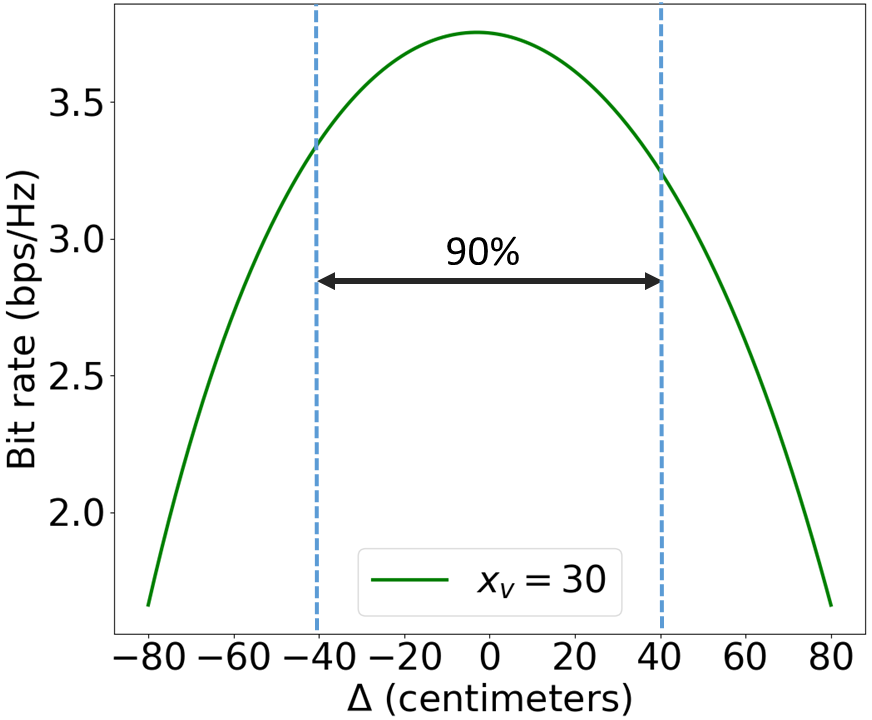}} 
    \subfloat[]{\includegraphics[scale=0.19]{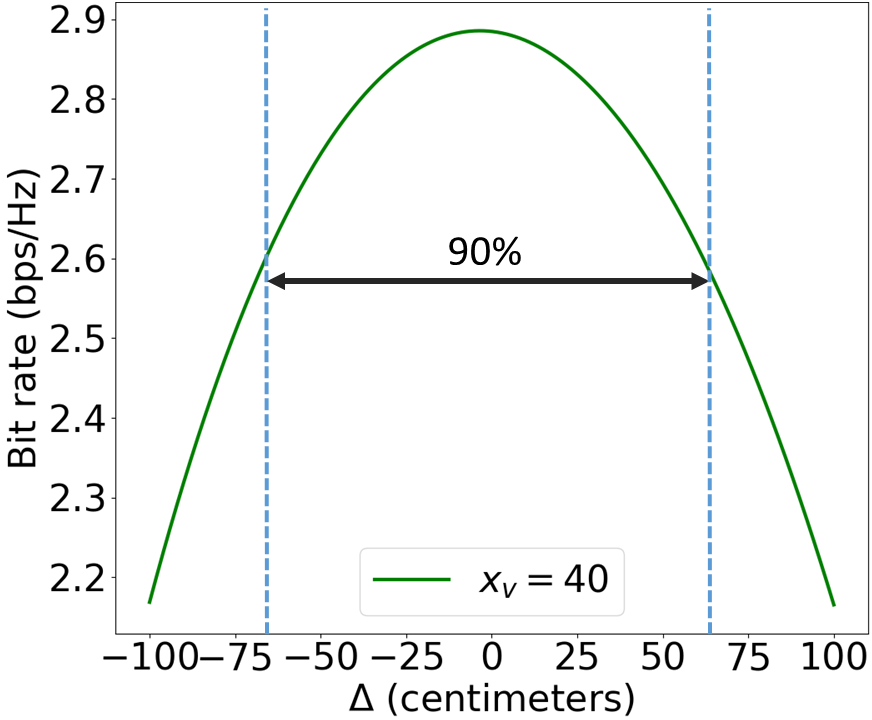}} 
    \subfloat[]{\includegraphics[scale=0.19]{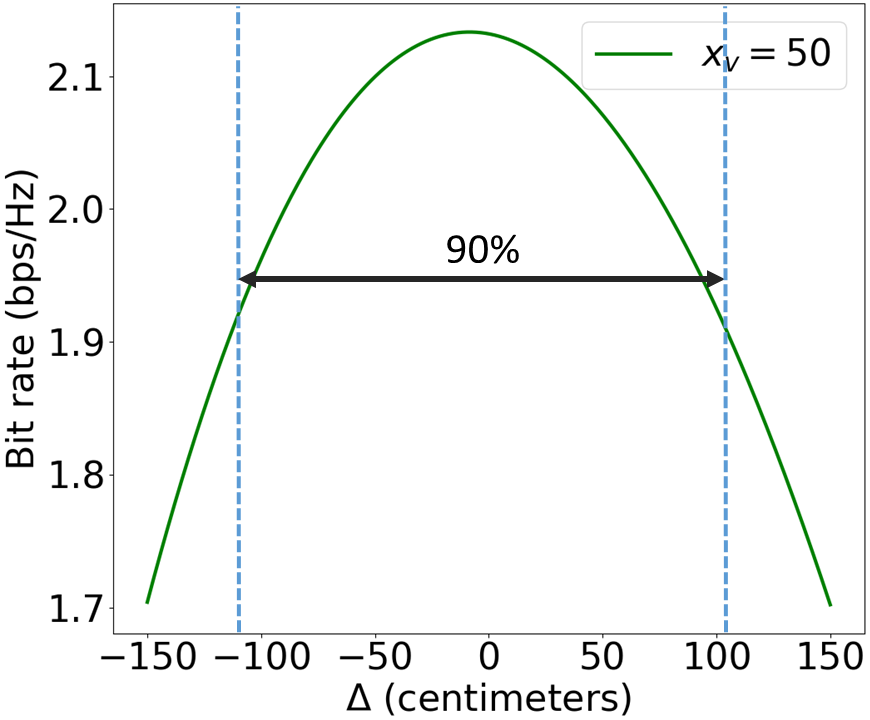}} 
    \caption{Inaccurate vehicle positioning effects on the bit rate at different positions ((a) $x_v=20$ (b) $x_v=30$ (c) $x_v=40$ (d) $x_v=50$).}
    \label{fig:time-slot-length}
\end{figure*}

In order to show the impact of that error $\Delta$, an experiment is conducted to realise how the bit rate is changed while varying the value of $\Delta$ with various vehicle positions (the distance from the vehicle to the RIS). The results are displayed in Fig. \ref{fig:time-slot-length} and indicate that the RIS can keep up to 90\% of its performance with error $\Delta$ ranging form 20 to 100 centimeters depending on the vehicle position. It is observed that distant vehicles from the RIS are less impacted by $\Delta$ as $\phi_{I,v}^n$ is less influenced by $\Delta$ when the distance between the vehicle and RIS is larger. One can also see in this figure that as $\Delta$ grows up, the bit rate decreases. Yet, the RIS elements can be tuned with plausible $\Delta$ values to maintain most of the original performance expected from RIS deployment.
According to \cite{humphreys2020deep}, vehicular  carrier-phase differential Global Navigation Satellite System (GNSS) positioning can estimate vehicle positions with accuracy of less than 17 centimeters accompanied by a success rate reaches up to 95\%. As shown in Fig.  \ref{fig:time-slot-length}, the minimum tolerance (to achieve 90\% of the performance) of vehicle positioning is 20 centimeters which is compatible with GNSS precision. Note that, for simplicity, in this work, $x_v^n$ is assumed to be accurately estimated.

%To sum up, thanks to high definition maps and simultaneous positioning techniques, it becomes possible to determine vehicle positions with centimeter-level accuracy\cite{zhang2019mobile, ma2019generation}. Furthermore, according to \cite{humphreys2020deep}, vehicular  carrier-phase differential Global Navigation  Satellite System positioning can estimate vehicle positions with accuracy of less than 17 centimeters and success rate reaches up to 95\%. 

\section{Simulation and Evaluation}
\label{sec:numerical-result}
\subsection{Simulation Setup}
As mentioned earlier, we use SUMO to mimic vehicular environment. Two flows of vehicles are generated; one with normal speed (max speed 50Kph) and the other with slow speed (max speed 30Kph). For the Deep Reinforcement Learning, 3 linear layers are used with tanh as activation function for the middle layers and softmax for the output layer. Internal layers contain 64 units each and Adam optimizer is incorporated to minimize the loss function. Learning rate is set to 0.002, $\gamma$ to 0.08, and clip to 0.02. The results were averaged over 500 tests. The remaining parameters used in our study are listed in Table \ref{table:expermint-parameters} (unless otherwise indicated).

\begin{table}[htbp]\small
	\caption{Simulation Parameters}
	\begin{center}
		\begin{tabular}{|c|c|}
		    \hline
		    \rowcolor{lightgray} Parameter& Value \\
		    \hline
		    Road segment length& 100 m\\
		    \hline
		    Arrival rate& 0.2 Veh/sec\\
		    \hline
		    $\sigma^2$ & $- 110$ dBm \\
			\hline
			$K$& 10 dB \cite{abdullah2020hybrid}  \\
			\hline
			$\alpha$& 4\\
			\hline
			$P$& 20 dBm \\
			\hline
			$C$& 3 \\
			\hline
			$\rho$ & 10 dBm\\
			\hline
			$M$ & 100 \\
			\hline
			$b$ & 2 bits \\
			\hline
			$x_I, y_I, z_I$& 10, 20, 10 \\
			\hline
			$x_R, y_R, z_R$& 0, 40, 10 \\
			\hline
			$y_v^n, z_v^n$& 20, 1 \\
			\hline
			\end{tabular}
		\label{table:expermint-parameters}
	\end{center}
    \end{table}

\subsection{Numerical Results}

\begin{figure}
	\centerline{\includegraphics[scale=0.25]{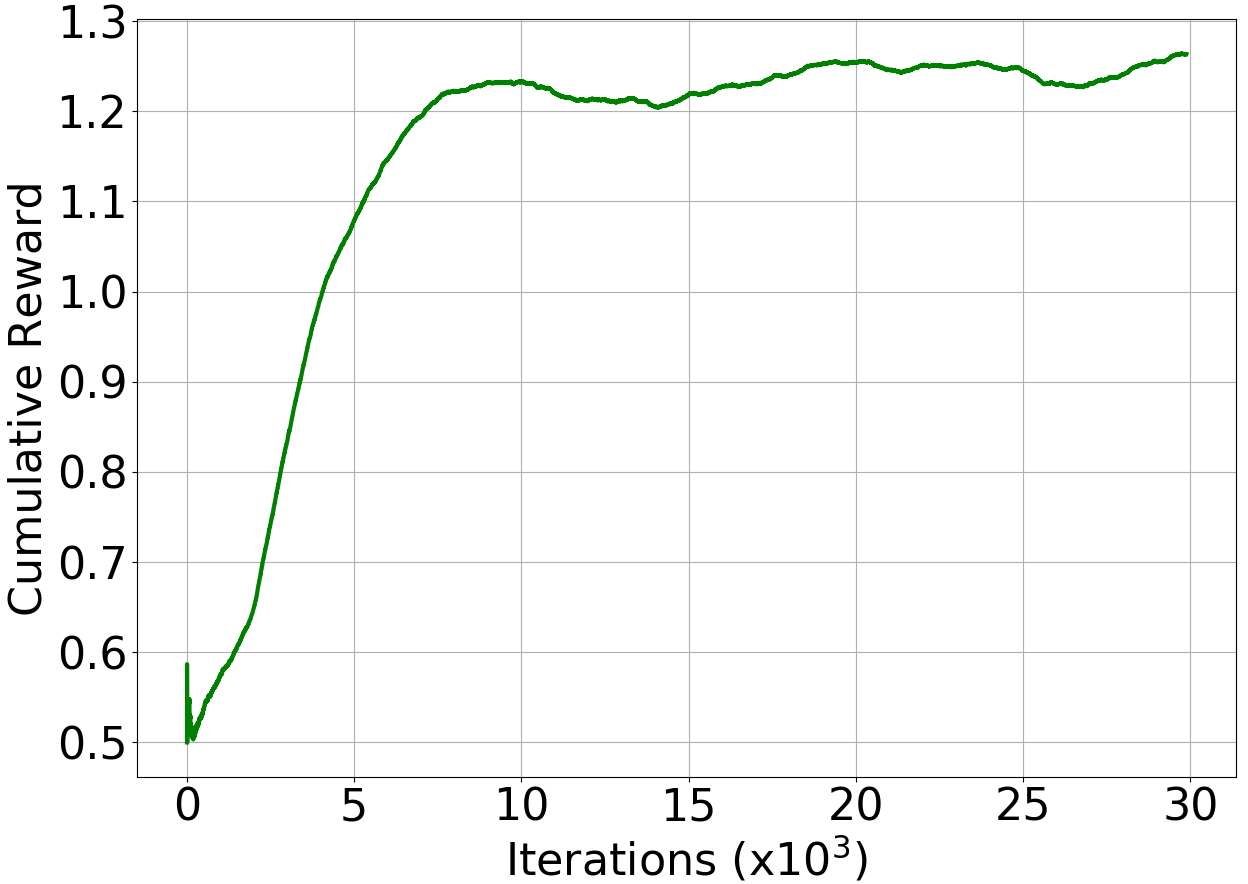}}
	\caption{Convergence over time.}
	\label{fig:result_convergence}
\end{figure}

First, we attempt to see the behaviour of the DRL agent. As illustrated in Fig. \ref{fig:result_convergence}, the cumulative reward, here represents minimum average bit rate, is remarkably increasing as the agent is exposed to more epochs/iterations. One can note that after around 7000 iterations, the system starts to converge.

In order to validate the performance of the proposed algorithm, it is compared  with  three  other benchmarks as follows
%Next, we will compare our solution with three benchmarks that are explained below.
\begin{itemize}
    \item \textbf{Greedy Scheduling with BCD (GS-BCD)}: In this scheme, the vehicle schedule sub-problem is solved with greedy algorithm; meanwhile the passive beamforming sub-problem is obtained using the proposed BCD scheme. The greedy algorithm can be explained as follows. At each time slot $n \in N$, a greedy algorithm ranks the set of vehicles $V^n$ based on their cumulative bit rate achieved up to $n$ ($z_v^n$). Then, those with the lowest average bit rates will be scheduled to be served in the following time slot.
    
    \item \textbf{Random Scheduling with BCD (RS-BCD)}: In this scheme, each time slot, the vehicles are randomly scheduled. While the proposed BCD is used for obtaining the passive beamforming at the RIS.
    
    \item \textbf{DRL with Random Phase-Shift Matrix (DRL-RPS)}:The proposed DRL algorithm is used to schedule RSU resources without any optimization over phase-shift for the RIS elements (Random values for the RIS elements' phase-shift).
\end{itemize}

It is worthwhile to compare with these baseline methods as they will show us how the performance would be if one of the two sub-problems are solved via an alternative widely-used method such as greedy or random while the second one is solved with the same method we propose.

\begin{figure}
	\centerline{\includegraphics[scale=0.25]{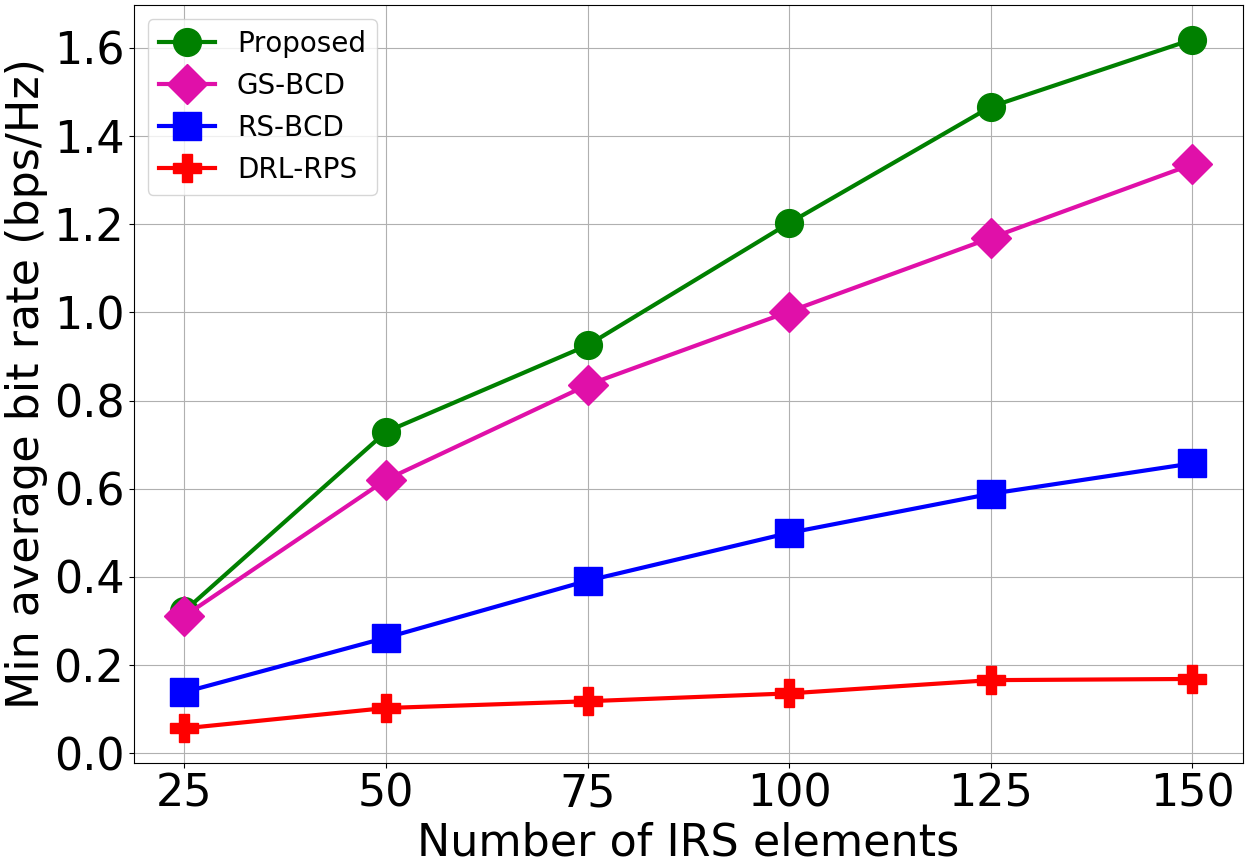}}
	\caption{RIS number of elements $M$ effects on network performance.}
	\label{results_no_elements}
\end{figure}

The effect of RIS number of elements $M$ is first studied. As demonstrated in Fig. \ref{results_no_elements}, with small number of elements, the achievable minimum average bit rate is very limited. However, as more elements are incorporated, the gain starts to grow up gradually, especially for our proposed solution approach. Another insight one can notice is the gap between the proposed solution with other methods over different $M$ values. It is very apparent that this gap widens proportionally as $M$ increases. With $M=150$, the difference between the proposed one and the greedy algorithm is about 17\%. Meanwhile, GS-BCD seems to have also a good performance compared to the other two methods. The reason behind that is GS-BCD attempts to reduce the minimum average bit rates for those vehicles with low bit rate levels. In addition, as GS-BCD leverages BCD, it also benefits from the good RIS configuration. RS-BCD, on the other hand, comes in the third place with clear gap from GS-BCD as it does not take into account low bit rate vehicles. At the end, DRL-RPS attains very poor performance that indicates that without a proper RIS configuration, the performance would be very poor even if the wireless scheduling is done carefully. 

\begin{figure}
	\centerline{\includegraphics[scale=0.25]{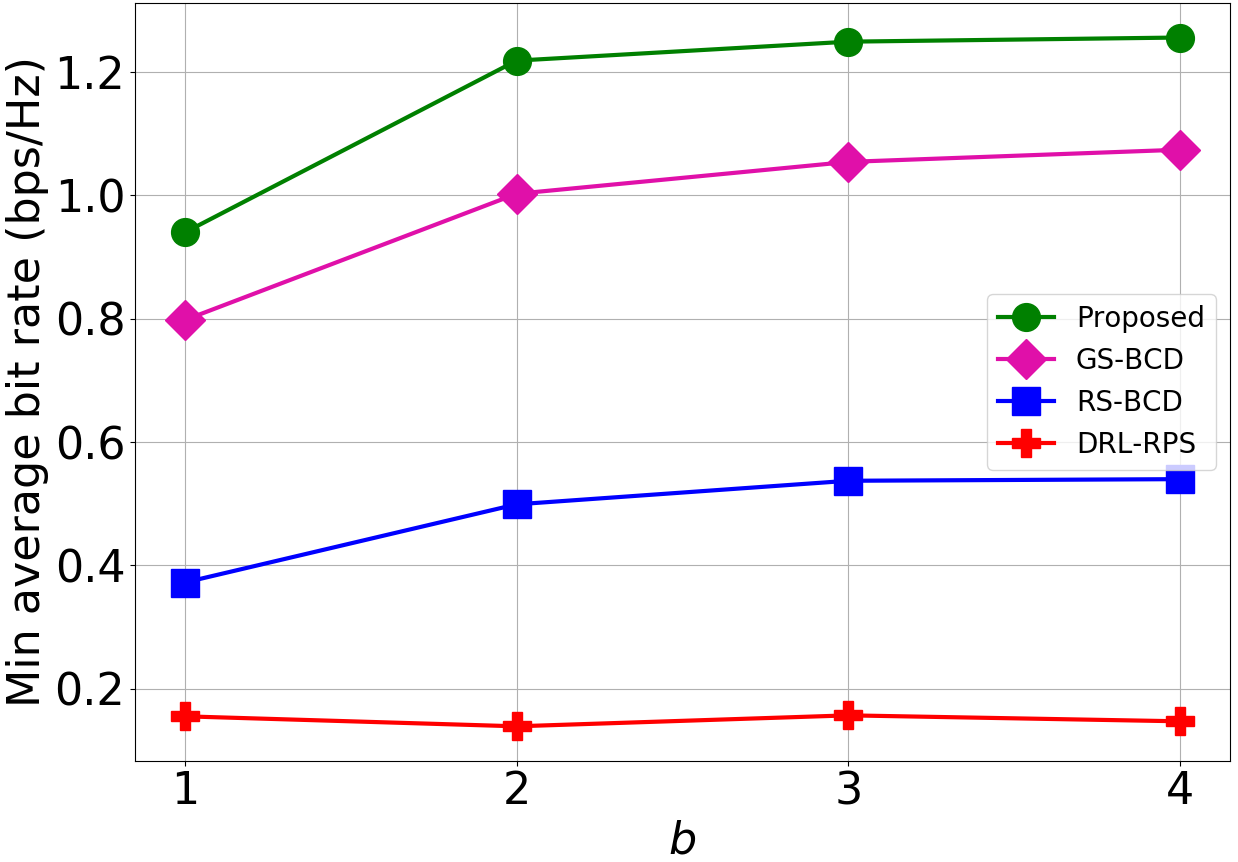}}
	\caption{Discrete quantization levels effects.}
	\label{results_control_bit}
\end{figure}

In our next experiment, we vary the value of $b$ for practical RIS. When $b$ is high, more phases are available for the configuration which is better for optimal RIS tuning. As it can be seen in Fig. \ref{results_control_bit}, larger $b$ means better performance. Yet, one can also notice that $b$ of 2 or 3 can almost obtain the highest gain. These behaviours have also been highlighted in other works related to non-dynamic environment \cite{di2020hybrid}. In the same figure, we can see that the proposed solution always achieves the highest performance regardless of $b$. Indeed, the difference can reach up to 19\% from GS-BCD. In the meantime. RS-BCD could only achieve minimal gain with larger $b$ and still below 0.6 of minimum average bit rate in all scenarios. In contrast to the other methods, DRL-RPS was unable to add any gain with larger $b$. That is due to the fact that this method does not consider RIS element tuning in the first place. Hence, higher $b$ values may also mean higher probability of falling to align with the vehicles as the options of angles for the RIS elements become larger.

\begin{figure}
	\centerline{\includegraphics[scale=0.25]{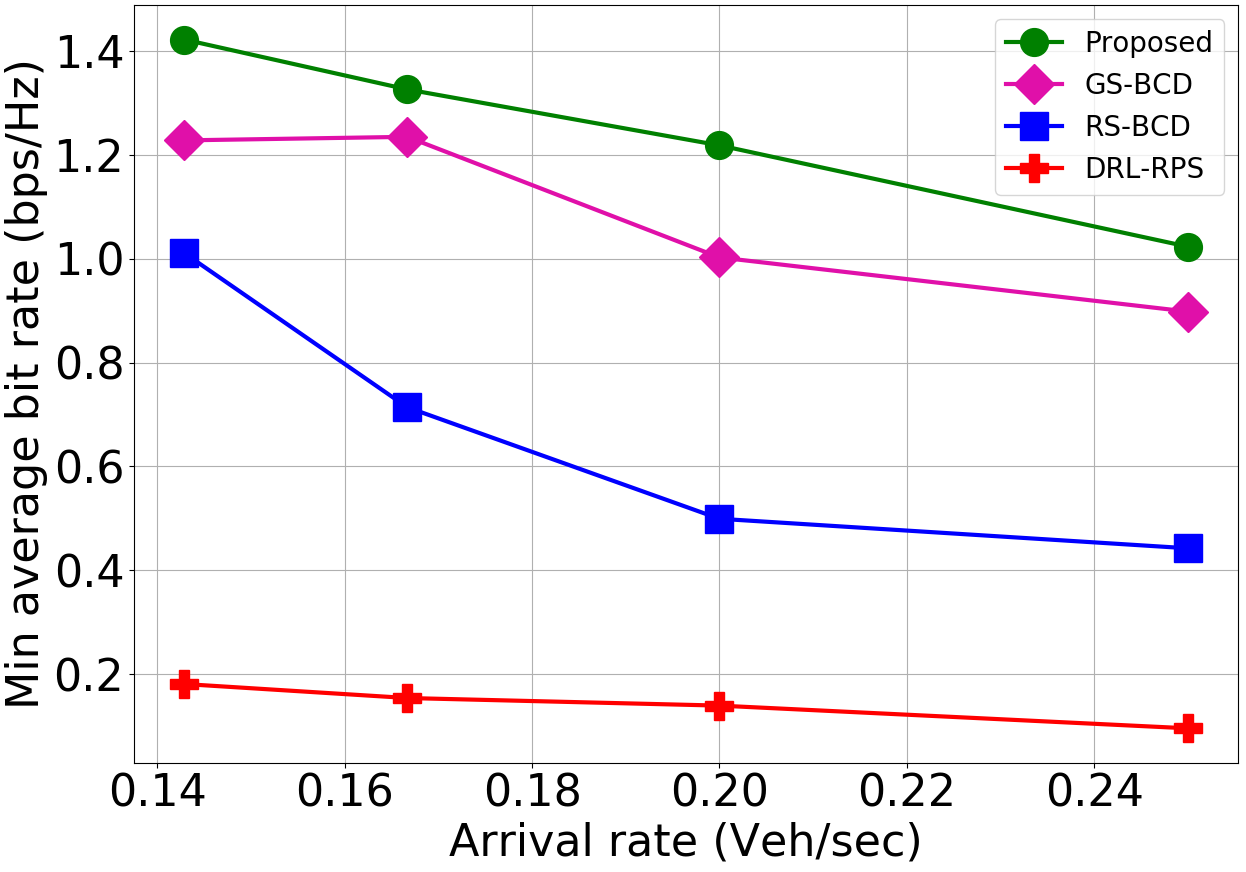}}
	\caption{Min average bit rate values over different vehicle arrival rates.}
	\label{fig:results_density}
\end{figure}

Next, the impacts of road density on the network is studied. Here, we vary the arrival rates of vehicles which results in more or less vehicles available simultaneously within the road segment. Intuitively, with small arrival rate, the RSU and RIS can better serve the vehicles. As seen in Fig. \ref{fig:results_density}, minimum average bit rate is slightly above 1.4 bps/Hz. However, as the road segment becomes more dense, the value degrades to approach approximately 1 bps/Hz. For the other methods, we can observe similar behaviours expect for GS-BCD where its gain seems to saturate at very low arrival rates. That is because GS-BCD does not consider the distance between the RIS and vehicles. It always assigns the resources for those with less $z_v^n$ regardless of their location or speed. Therefore, the wireless resources might be wasted on far vehicles instead of making benefit by serving nearest ones. Also, selecting two or more vehicles with relatively large gap between them reduces the efficiency of the RIS to serve both of them since the RIS needs to maximize the sum of immediate bit rates which is undesirable in such scenario\footnote{Note that, the immediate average bit rate, denoted by $z_v^n$, does not necessarily reflect the ultimate average bit rate of vehicle $v$. It only represents what was the average bit rate up to $n$ which depends on the time elapsed since vehicle $v$ arrival. Intuitively, this value decreases over time as the elapsed time increases, and this issue is not taken into consideration by GS-BCD.}. This especially appears when less numbers of vehicles exist on the road and the RSU oftentimes schedules for distant vehicles. In contrast, RS-BCD attains steep increase in minimum average bit rates with low road density since there will be less vehicles and the probability of vehicles being served is much higher. This impact is much less significance with DRL-RPS due to the miss alignment in phase-shifts with the vehicles.

Finally, one of the insightful indices to use in similar problem of max-min is Jain's fairness index \cite{sediq2013optimal}. The formula for this index is:
\begin{equation}
    \dfrac{(\sum_{v=1}^{V} z_v)^2}{V \sum_{v=1}^{V} (z_v)^2} 
\end{equation}

\begin{figure}
	\centerline{\includegraphics[scale=0.25]{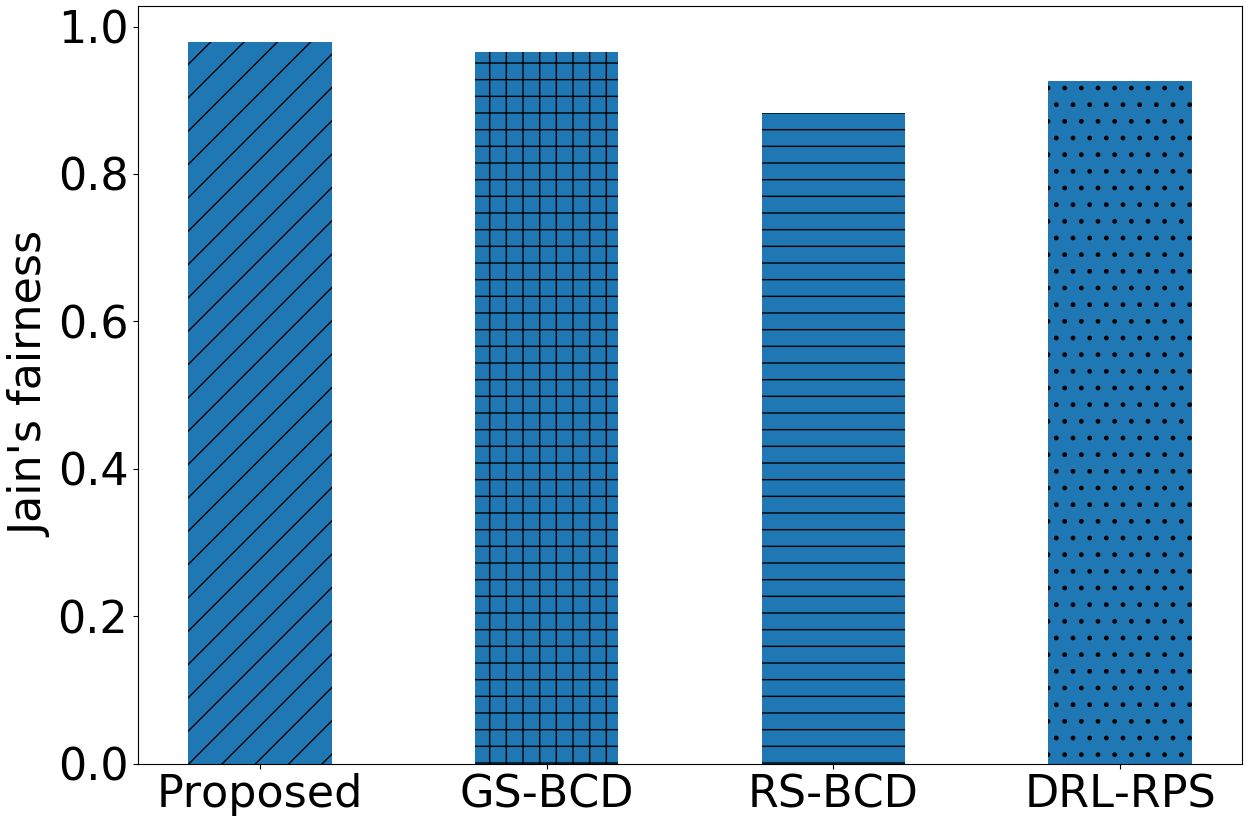}}
	\caption{Comparison for the Jain\'s fairness ($M=100$).}
	\label{fig:results_jain}
\end{figure}

Then, we conduct an experiment to see the levels of jain's fairness attained by the four algorithms. We can notice in Fig. \ref{fig:results_jain} that our proposed solution achieved the highest level of fairness in comparison to the other methods. But, one can also see that the four methods, in fact, obtain high levels in general. For GS-BCD, it actually attempts to reduce the discrepancies in minimum average bit rates among all vehicles, hence, it enhances the fairness. RS-BCD, on the other hand, schedules the resources at random with uniform distribution which also improves the fairness. Finally, DRL-RPS leverages our DRL agent which indeed tries to maximize the original maxmin problem. That is, all the methods are able to maintain nice levels of fairness among the vehicles. Despite this fact, it is still true that only our proposed solution approach can achieve the highest fairness levels while notably maximizing the minimum average bit rates through considering the coupling effects of the two sup-problems.

\section{Conclusion}
\label{sec:conclusion}
We have investigated the area of RIS integration with vehicular communications. That is, the core of this work evolved around a system model that employs RIS to provide favourable wireless experiences for vehicles travelling in a dark zone. The RIS has demonstrated high competence in establishing indirect links between the RSU and vehicles. Throughout this study, we have also seen that DRL is an appealing solution to cope with the highly dynamic nature of such environment and it can adapt to various road conditions and RIS options. In addition, BCD was also leveraged to provide efficient yet robust solutions to the RIS phase-shift matrix. In the numerical results, the performance of our solution method has been analyzed thoroughly by comparing it with other benchmarks. Aside from that, we have also carried out a study on RIS placement to attain optimized wireless communication with the RSU and non-static receivers.

In terms of future work, we will extend this work considering wireless resource allocation where the spectrum can be allocated to each vehicle based on its individual needs. As a result, the RIS phase-shift configuration method can also be updated to consider various link qualities for each specific vehicle depending on the allocated wireless resources.  

%we will investigate the roles that Unmanned Ariel Vehicle (UAV) equipped with RIS can play to further extend and improve the coverage of the RSU. Given their flexibility and deployability, UAVs can better adapt to the dynamic nature of vehicular environment than fixed IRS deployment.  

\bibliographystyle{IEEEtran}
\bibliography{IEEEabrv, references}
\end{document}